\documentclass[aps,prl,reprint,superscriptaddress,amsmath,amssymb,floatfix]{revtex4-2}

\usepackage{graphicx}
\usepackage{bm}
\usepackage{siunitx}
\usepackage{booktabs}
\usepackage{xcolor}
\usepackage{microtype} % fewer under/overfull lines
\usepackage{xurl} % allow line breaks anywhere in long URLs
\usepackage[colorlinks=true,allcolors=blue]{hyperref}
\usepackage[nowarn]{glossaries}
\glsdisablehyper
\newacronym{mlip}{MLIP}{machine-learning interatomic potential}
\newacronym{aimd}{AIMD}{\textit{ab initio} molecular dynamics}
\newacronym{md}{MD}{molecular dynamics}
\newacronym{dft}{DFT}{density-functional theory}
\newacronym{film}{FiLM}{feature-wise linear modulation}
\newacronym{grace}{GRACE}{graph atomic cluster expansion}
\newacronym{lorem}{LOREM}{learning long-range representations with equivariant messages}
\newacronym{razor}{RAZOR}{response analysis in $z$-orientation}
\newacronym{lammps}{LAMMPS}{Large-scale Atomic/Molecular Massively Parallel Simulator}

\newcommand{\lorem}{\glsentryshort{lorem}}
\newcommand{\grace}{\glsentryshort{grace}}
\newcommand{\razor}{\glsentryshort{razor}}

\newcommand{\graceq}{\grace-$q$}
\newcommand{\loremq}{\lorem-$q$}

\newcommand{\prlsec}[1]{\textit{#1.}---}

\makeatletter
\newcommand{\nocontentsline}{\let\origaddcontentsline\addcontentsline\let\addcontentsline\@gobblethree}
\newcommand{\restorecontentsline}{\let\addcontentsline\origaddcontentsline}
\makeatother

\usepackage[normalem]{ulem}
\newcommand{\PL} [1] {{\color{magenta} #1}}

\begin{document}

\title{Bias Conditioning and Long-Range Electrostatics for Consistent Machine-Learning Potentials of Electrified Interfaces}

\author{Lucas B. T. de Kam}
\email{lucas.de.kam@tum.de}
\affiliation{Department of Chemistry, School of Natural Sciences, Technical University of Munich, Garching, Germany}
\affiliation{Atomistic Modeling Center, Munich Data Science Institute, Technical University of Munich, Garching, Germany}
\author{Patrick Rinke}
\email{patrick.rinke@tum.de}
\affiliation{Atomistic Modeling Center, Munich Data Science Institute, Technical University of Munich, Garching, Germany}
\affiliation{Department of Physics, School of Natural Sciences, Technical University of Munich, Garching, Germany}
\affiliation{Munich Center for Machine Learning (MCML)}
\author{Christopher J. Stein}
\email{christopher.stein@tum.de}
\affiliation{Department of Chemistry, School of Natural Sciences, Technical University of Munich, Garching, Germany}
\affiliation{Atomistic Modeling Center, Munich Data Science Institute, Technical University of Munich, Garching, Germany}
\author{Philip Loche}
\email{philip.loche@tum.de}
\affiliation{Atomistic Modeling Center, Munich Data Science Institute, Technical University of Munich, Garching, Germany}
\affiliation{Department of Physics, School of Natural Sciences, Technical University of Munich, Garching, Germany}
\affiliation{Munich Center for Machine Learning (MCML)}

\date{\today}

% ---------------------------------------------------------------------------
% Abstract (600 characters)
% ---------------------------------------------------------------------------
\begin{abstract}
\Glspl{mlip} of electrified interfaces need the bias as input, but it is unclear whether a scalar electrode charge suffices and whether short-range models capture the non-local response.
We add charge and field conditioning to short- and long-range \glspl{mlip}.
State-of-the-art field-conditioned short-range models overpolarize thick water films nearly twofold and miss work-function shifts at stepped Pt.
Our long-range model captures both, needs only the electrode charge at metals as input, and reproduces \textit{ab initio} simulations of the Pt(111)--water interface.
\end{abstract}

\maketitle
\glsresetall  % abstract and main text define acronyms independently

% ---------------------------------------------------------------------------
% Main text (PRL: 3750 words incl. captions; ~4 figures)
% ---------------------------------------------------------------------------

% --- Introduction (no run-in heading, as in PRL) ---
A key requirement for modeling electrochemical interfaces is the ability to apply an electrical bias.
The applied bias controls the structure and dynamics of the interfacial electrolyte \citep{mathanker2026estimating}, the charge distribution at the electrode \citep{li_electron_2025}, and the rate of electron transfer \citep{zhang2026electrochemical}.
\textit{Ab initio} simulations impose this bias either through an excess electrode charge balanced by a countercharge (jellium, a charged plate, or a layer of inert atoms)~\cite{surendralal2018first, kastlunger_controlled-potential_2018,zhou_abacus_2025,mattoso_user-defined_2026} or through an applied field with a dipole correction~\citep{stengel_electric_2009}.
Both routes fix the same boundary condition, a displacement field $\mathbf{D}$ normal to the surface~\cite{zhang_finite_2016,brezina_electrochemical_2026}, and thus sample the constant-charge ensemble.
To simulate at constant potential instead of constant charge, a potentiostat adjusts the charge over time so that the electrode potential fluctuates around a target value~\cite{bonnet2012first,deisenbeck_dielectric_2023,sechi2025first,sun_electrochemical_2026}.
In either ensemble, however, \gls{aimd} is too expensive to converge even the potential of zero charge~\cite{hagopian_assessing_2026}. \Glspl{mlip} remove this cost barrier and are now being extended to biased interfaces. % maybe also/instead mention that long sampling times are required for (enhanced) sampling of reactions for example / higher levels of theory / more realistic, larger systems
To serve both ensembles, a biased \gls{mlip} must take the electrode charge, i.e., $\mathbf{D}$, as input and return the electrode potential, the charge derivative of its energy, as output.
%
% Lucas's previous version (kept for reference):
% The input and output requirements of a biased MLIP are determined by how the bias is controlled in molecular dynamics (MD).
% The control of a bias is usually done either by constant-charge or constant-displacement-field simulations require the MLIP to take the charge bias as an input.
% Constant-potential simulations need a potentiostat algorithm to sample the correct dynamics~\cite{bonnet2012first, deisenbeck_dielectric_2023, sechi2025first, sun_electrochemical_2026}.
% A potentiostat varies the charge so that the potential fluctuates about a controlled mean value, and thus also requires the MLIP to take charge as an input and potential as an output.
% Here, we therefore focus on designing MLIPs as constant-displacement-field-models, without loss of general applicability.

Existing \gls{mlip} designs differ in how the bias enters the model \citep{aouichaoui_machine_2026}.
A common approach is to embed the electrode charge or electron number as a scalar~\cite{wang_constant-potential_2025,sun_probing_2025,chen_constant-potential_2026,sahel-schackis_equifilm_2026}, following the total-charge inputs established for molecules~\cite{unke_spookynet_2021,xie_incorporating_2020,ko_fourth-generation_2021,batatia_mace-polar-1_2026}.
Field-aware models instead couple an applied vector field to equivariant features~\cite{gao_self-consistent_2022,joll_machine_2024,falletta_unified_2025,martin_general_2026,bergmann_machine_2025}.
% Each route commits to either a scalar or a vector input.
% A scalar electrode charge carries no direction and therefore cannot fix the sign of the response when both sides of the cell are geometrically equivalent.
Whether a scalar input suffices or the field direction is required has not been established.
Independent of the input, the response to a bias is non-local. %Crucially, ... also long-ranged.
The electrode potential or work function depends on the orientation of every solvent molecule in the cell, not only of those close to the surface. 
The solvent orientation, in turn, is limited by its own depolarizing field. 
Short-range \glspl{mlip}, including most of the models above, miss such long-range contributions: they overpolarize water~\cite{gao_self-consistent_2022,parker_false_2026,de_kam_modeling_2026}, %can behave as false metals and 
but the resulting errors remain hidden in energy and force metrics and appear only in sampled observables~\cite{sumic_electrostatic_2026,yang_macro-dipole-constrained_2026,grasselli_long-range_2026}.
The few long-range \glspl{mlip} for biased interfaces~\cite{zhu_machine_2025,wang_ion-modulated_2026} learn only the electrolyte charges and treat the metal with classical Gaussian charges~\cite{siepmann_influence_1995}, which miss the electron spillover into the interfacial water found in \textit{ab initio} simulations~\cite{li_electron_2025}; learning the electronic response of the metal instead predicts the polarization more accurately~\cite{grisafi_predicting_2023}.
Two questions therefore remain open: in which form should the bias enter an \gls{mlip}, and which interaction range does a consistent response require?

In this Letter, we address both questions by implementing charge and field conditioning in the short-range \gls{grace}~\cite{bochkarev_graph_2024} backbone and in our architecture for \gls{lorem}~\cite{rumiantsev_learning_2026}.
\lorem\ learns latent charges and feeds their Ewald-summed Coulomb potentials back into the descriptor features as long-range messages, so that the predicted properties depend on the electrostatic potential of the entire cell.
To show that long-range interactions are required for a correct bias response, we first consider water between two inert electrodes, a system that we can validate against theoretical predictions.
We then turn to metal--water interfaces and compare scalar and vector conditioning in short- and long-range models for a single water molecule above flat and stepped Pt surfaces.
Finally, constant-charge \gls{md} of the Pt(111)--water interface yields the bias-dependent water structure and electrode potential. 
Together, these systems show that long-range electrostatics is required to describe the bias response beyond the cutoff radius, and that a long-range model needs only a scalar input at metal--water interfaces.

\begin{figure*}
  \centering
  \includegraphics{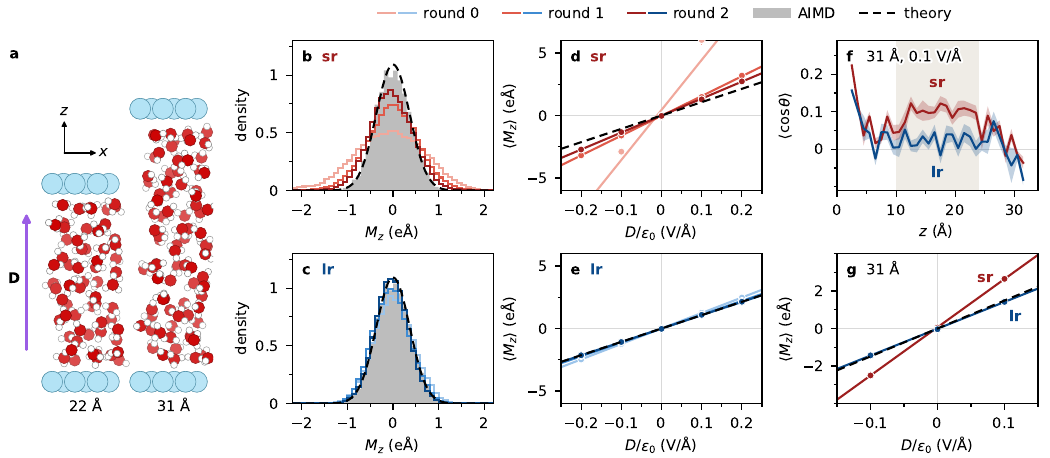}
  \caption{\Gls{md} of water between Ne electrodes at fixed displacement field. (a) Cells with the labeled \qty{22}{\angstrom} and the unseen \qty{31}{\angstrom} water film. (b,c) Distribution of $M_z$ at $D =\varepsilon_0E_0 = 0$ for the short-range (sr) and long-range (lr) models over three active-learning rounds, against \gls{aimd} (gray) and a Gaussian predicted by continuum electrostatics (black, dashed). (d,e) $\langle M_z\rangle$ from molecular dynamics at $E_0=D/\varepsilon_0=0,\pm0.1,\pm0.2$ \unit{\volt/\angstrom}; dashed line: expected response based on the \gls{aimd} fluctuations, Eq.~\eqref{eq:fluctuation}. (f,g) The round-2 models for the \qty{31}{\angstrom} film, which is not in the training set: (f) water orientation $\langle\cos\theta\rangle(z)$ with $\theta$ the angle between the water bisector and the $z$-axis, at a field of $D/\varepsilon_0 = \qty{0.1}{\volt\per\angstrom}$; the shading marks the water at least \qty{10}{\angstrom} from both Ne layers; (g) $\langle M_z\rangle$ against $D/\varepsilon_0$; dashed line: the continuum value $\varepsilon_0 A L_\mathrm{w}(1-1/\varepsilon)$. Error bars and bands are block estimates.}
  \label{fig:dipole}
\end{figure*}

\prlsec{Long-range interactions under bias} A capacitor consisting of water between two inert electrodes is a simple model system to probe the effect of long-range electrostatics under bias.
The bias can be introduced by charging the electrodes with opposite charges $\pm q$, which fixes the displacement field at $\mathbf{D}=(q/A)\hat{\mathbf{z}}$, where $A$ is the lateral area and $\hat{\mathbf{z}}$ the surface normal.
Here, we realize the same boundary condition by applying an electric field $E_0\hat{\mathbf{z}}$ together with a dipole correction, so that $\mathbf{D}=\varepsilon_0 E_0\hat{\mathbf{z}}$. Note that $E_0$ is the field in the empty cell; the macroscopic field inside the water, $\mathbf{E}=(\mathbf{D}-\mathbf{P})/\varepsilon_0$, is screened by the water polarization $\mathbf{P}$. As shown in Sec.~\ref{sm:newaterne-energy} of the Supplemental Material, the energy of this system can be written as 
\begin{equation}
\begin{aligned} \label{eq:capacitor}
U(\mathbf{D}) = U_0 - \frac{1}{\varepsilon_0} \mathbf{M} \cdot \mathbf{D} = U_0 - M_zE_0,
\end{aligned}
\end{equation}
where $U_0$ is the energy at zero displacement field and $\mathbf{M}$ the cell dipole moment. The bias response of the energy is thus given by the cell dipole in $z$, $M_z=-\partial U/\partial E_0$.

The capacitor is symmetric under $z\to-z$, so its geometry does not reveal the field direction. An \gls{mlip} must therefore receive this direction explicitly to learn the bias-dependent energy. We couple the applied field $E_0\hat{\mathbf{z}}$ to the equivariant features of the model, similar to Ref.~\citenum{martin_general_2026} (see Sec.~\ref{sm:field} of the Supplemental Material for the implementation). With $M_z$ obtained by automatic differentiation of $U=\sum_i U_i$, the model represents the cell dipole as a sum of atomic dipoles, $M_z=\sum_i \mu_{i,z}$.

To isolate the effect of long-range electrostatics, we implement field conditioning in \lorem{} and compare its long-range and short-range variants. The long-range model uses an Ewald sum with latent charges; the short-range model has a \qty{5}{\angstrom} cutoff with one message-passing step, leaving it with an effective cutoff of at most \qty{10}{\angstrom}. We train these models on the energy $U$, forces $\mathbf{F}_i$, and cell dipole $M_z$ as learning targets. The training set consists of Ne--water--Ne structures with a \qty{22}{\angstrom} water film (Fig.~\ref{fig:dipole}a). We sample frames from a zero-field \gls{aimd} trajectory from Ref.~\citenum{yang_macro-dipole-constrained_2026}, label them using single-point calculations at $|E_0|\leq \qty{0.2}{\volt\per\angstrom}$, and add two rounds of active learning. Each round adds a few hundred structures sampled from \gls{md} trajectories driven by the models of the previous round (see Sec.~\ref{sm:ne-data} of the Supplemental Material for dataset and training details).

In zero-field \gls{md}, the short-range models show substantially larger fluctuations in $M_z$ than the long-range models (Fig.~\ref{fig:dipole}b,c), as reported previously~\citep{gao_self-consistent_2022, parker_false_2026, de_kam_modeling_2026}. Active learning improves the agreement of the short-range models with \gls{aimd}, but they never reach full agreement for the training set sizes tested here. In contrast, the long-range models reproduce the \gls{aimd} result with little additional data. 

The field response of $M_z$ is related to the variance of its fluctuations (see Sec.~\ref{sm:fluctuations} of the Supplemental Material for the derivation),
\begin{equation} \label{eq:fluctuation}
    \frac{\partial \langle M_z\rangle}{\partial E_0} \bigg|_{E_0=0} = \bigg\langle \frac{\partial M_z}{\partial E_0}\bigg|_{\mathbf{R}}\bigg\rangle + \beta \text{Var}(M_z),
\end{equation}
where $\partial M_z/\partial E_0|_\mathbf{R}$ is the electronic response at fixed nuclear positions. Evaluating this first term by finite differences on a few snapshots labeled at $E_0=\pm\qty{0.1}{\volt\per\angstrom}$ and obtaining the variance from the zero-field \gls{aimd} trajectory yields a reference for $\partial \langle M_z\rangle/\partial E_0|_{E_0=0}$. Figure~\ref{fig:dipole}d,e compares this slope (dashed black line) with finite-field \gls{md}. The short-range models deviate from the reference even after two active-learning rounds, consistent with their larger variance in Fig.~\ref{fig:dipole}b,c. The long-range model, in contrast, matches the reference after a single round.

The necessity of long-range interactions becomes even more apparent when extrapolating to a thicker, unseen \qty{31}{\angstrom} water film (Fig.~\ref{fig:dipole}a, right).
Figure~\ref{fig:dipole}f shows the water orientation at \qty{0.1}{\volt\per\angstrom} in the trajectories obtained with the short-range and long-range round-2 models. While the short-range model behaves similarly to the long-range model near the electrodes, it strongly overpolarizes the water in the center of the film, where the interfaces are outside the model's effective cutoff. The long-range model extrapolates well: in Fig.~\ref{fig:dipole}g, its response lies close to the reference (black dashed line), whereas the short-range model overestimates the response nearly twofold. In the absence of an \gls{aimd} trajectory at this film thickness, we estimate the reference from the \qty{22}{\angstrom} cell, adding the response $\varepsilon_0A\,\Delta L(1-1/\varepsilon)$ of the additional bulk water with thickness $\Delta L\approx\qty{9}{\angstrom}$ (see Sec.~\ref{sm:continuum} of the Supplemental Material for details). Field-conditioned long-range models thus correctly describe the polarization of water under bias, whereas field-conditioned short-range models fail to do so.

\begin{figure}
  \centering
  \includegraphics[width=\linewidth]{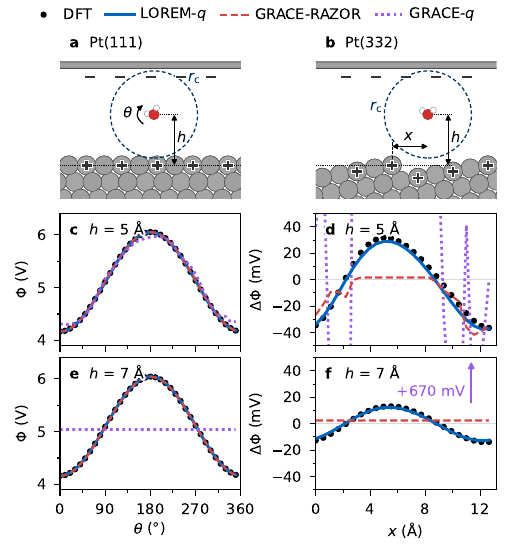}
  \caption{Work function of one water molecule above flat and stepped Pt. (a) Water at height $h$ above Pt(111), its dipole rotated by $\theta$; (b) water moved by $x$ across a step of Pt(332). Circles mark the cutoff $r_\mathrm{c}$. (c,e) $\Phi$ against $\theta$ on Pt(111) at $h=5$ and \qty{7}{\angstrom}. (d,f) $\Delta\Phi$, relative to the \textit{ab initio} mean, against $x$ on Pt(332); at $h=\qty{7}{\angstrom}$ \graceq\ lies off scale (\qty{+0.67}{\volt}).}
  \label{fig:pt}
\end{figure}

\begin{figure*}
  \centering
  \includegraphics[width=\textwidth]{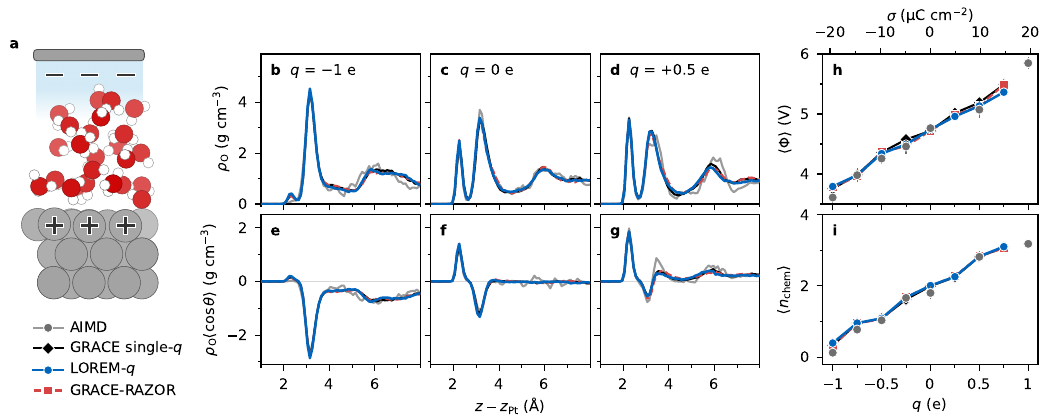}
  \caption{Constant-charge \gls{md} of Pt(111)--water. (a) Simulation snapshot of a $3\times4\times3$ Pt(111) slab with 24 water molecules and a solvated jellium countercharge \citep{kastlunger_controlled-potential_2018}. (b--d) Oxygen density $\rho_\mathrm{O}$ and (e--g) orientation profile $\rho_\mathrm{O}\langle\cos\theta\rangle$ against the height above the top Pt layer, at $q=-1$, $0$, and $+0.5\,e$. (h) Mean work function or electrode potential against $q$ (top axis: charge density $\sigma$); its slope is the inverse capacitance. The work function of \grace{} single-$q$ is evaluated with \loremq{} on its trajectories. (i) Mean number of chemisorbed water molecules $\langle n_\mathrm{chem}\rangle$, counted as O atoms less than \qty{2.6}{\angstrom} above the top Pt layer. }
  \label{fig:capacitance}
\end{figure*}

\prlsec{Conditioning for metal--water interfaces} The metal--water--countercharge setups used in computational electrocatalysis are essentially the same constant-$\mathbf{D}$ problem as the capacitor model considered above, with one inert electrode replaced by a metal slab. However, two things change. First, the bias charge sets the number of electrons, so the bias response is the work function $\Phi=\partial U/\partial q$, where $q$ is in units of the elementary charge $e$. Second, the electronic charge is free to move and can, for example, spill over into the electrolyte~\citep{li_electron_2025}.
The position of the interfacial charge therefore depends on the interfacial structure. As shown in Sec.~\ref{sm:workfunction} of the Supplemental Material, at any charge,
\begin{equation} \label{eq:work-function-decomposition}
    \Phi = \Phi_\mathrm{ref} - \frac{\Delta M_z}{\varepsilon_0 A} + \frac{qd}{\varepsilon_0 A},
\end{equation}
where $\Phi_\mathrm{ref}$ is the work function of a reference system such as the bare metal slab and $\Delta M_z = \sum_i \Delta \mu_{i,z}$ is the change in dipole moment relative to this reference. $d=\bar{z}_\mathrm{cc} - \bar{z}_q$ is the distance between the centroid $\bar{z}_\mathrm{cc}$ of the countercharge and the centroid $\bar{z}_q$ of the excess charge $q$ on the metal and water atoms. Note that $\Delta M_z$ contains the electronic rearrangement of the metal surface as well as that of the water. The potential energy surface is then given exactly by
\begin{equation} \label{eq:charging-integral}
    U(q) = U_0 + \int_0^q \Phi(q')\,\mathrm{d}q',
\end{equation}
with $U_0=U(q=0)$.
Assuming that $\Delta M_z$ depends only linearly on $q$ and that $d$ is constant, we recover the quadratic expansion~\citep{bergmann_machine_2025}
\begin{equation} \label{eq:razor-expansion}
    U(q) = U_0 + q\Phi_0 + \frac{q^2}{2C_0},
\end{equation}
with the zero-charge work function and capacitance
\begin{align}
    \Phi_0 = \Phi_\mathrm{ref} - \frac{\Delta M_{z,0}}{\varepsilon_0 A}, &&
    \frac{1}{C_0} = \frac{1}{\varepsilon_0 A}\left(d - \frac{\partial \Delta M_z}{\partial q}\right),
\end{align}
with $\Delta M_{z,0}=\Delta M_z(q=0)$.
Compared to the capacitor model from Eq.~\eqref{eq:capacitor}, the metal--water--countercharge system introduces a linear term $q\Phi_\mathrm{ref}$ and a quadratic term $q^2/(2C_0)$, both of which are properties of the data and must be learned by the \gls{mlip}. 
The assumptions of constant $d$ and linear $\Delta M_z(q)$ imply that the capacitance $C_0$ is constant.
This holds only in a narrow bias window around the value of $q$ at which the water film has been equilibrated~\citep{bergmann_erasing_2026}. 
Outside that window, the electron excess or deficit is deposited in states belonging to the water, so that $\bar{z}_q$ changes and $d$ is no longer constant.

Many existing strategies~\citep{sun_probing_2025, wang_constant-potential_2025, chen_constant-potential_2026} condition an $O(3)$-symmetric model on a scalar $q$ to learn $U(q)$. 
These models cannot resolve the field direction for water molecules with the surface outside their effective cutoff, and therefore cannot give an orientation-dependent contribution to the bias response.
While field conditioning can address the lack of directionality~\citep{bergmann_machine_2025}, the limitations of this approach near rough surfaces with an inhomogeneous field have not been tested.
We now investigate the limitations of short-range models using our own implementation in the \gls{grace} backbone, and show that the charge-conditioned long-range \gls{lorem} backbone overcomes them.
Previous charge-conditioning models concatenate $q$ to the descriptor features~\citep{sun_probing_2025, wang_constant-potential_2025}; we instead apply a \gls{film}~\citep{perez_film_2018} layer, a more general way to condition a neural network on a global variable. The work function $\Phi=\partial U/\partial q$ and the surface-normal components of the Born effective charges, $\mathbf{Z}^*_{i,z} = \varepsilon_0 A\, \partial \mathbf{F}_i/\partial q$, then follow by automatic differentiation. We denote the resulting models \graceq{} and \loremq{} (see Sec.~\ref{sm:charge} of the Supplemental Material for implementation details).
To complete the comparison, we use the field-conditioned \gls{razor} approach~\citep{bergmann_machine_2025}, which learns the zero-charge work function $\Phi_0$ of Eq.~\eqref{eq:razor-expansion} from latent atomic dipoles and the capacitance $C_0$ as a constant. While \gls{razor} originally used two backbones, our implementation, denoted \grace-\razor{}, uses a single \grace{} backbone (see Sec.~\ref{sm:razor} of the Supplemental Material for implementation details).

We train the three models (\grace{}-\razor{}, \graceq, and \loremq) on a single water molecule above a Pt surface. We consider two surfaces: flat Pt(111) (Fig.~\ref{fig:pt}a) and stepped Pt(332) (Fig.~\ref{fig:pt}b). To teach the models the charge response, the data are labeled at different charges in a small window around $q=0$ ($|\sigma|\leq\qty{0.5}{\micro\coulomb\per\centi\metre\squared}$), using a plate \qty{13}{\angstrom} above the surface as countercharge. We train with energy, forces, and work function as learning targets (see Sec.~\ref{sm:pt} of the Supplemental Material for dataset and training details).

For Pt(111), all models predict the correct work function when the water is inside the cutoff (Fig.~\ref{fig:pt}c). With the water beyond the cutoff (Fig.~\ref{fig:pt}e), the charge-conditioned \graceq{} model fails to reproduce the response. The field-conditioned \grace{}-\razor{} model still gives the correct response for the flat surface, even with the water outside the cutoff. Its field input carries the direction that the water molecule cannot infer from the geometry.
For the stepped surface (Fig.~\ref{fig:pt}d,f), however, \grace{}-\razor{} fails to reproduce the response when the water molecule is moved across the step in the $x$ direction. A homogeneous applied field cannot describe the locally inhomogeneous field near the step. At $h=\qty{5}{\angstrom}$, where the step changes $\Phi$ by about \qty{30}{\milli\volt}, the work-function error of \grace{}-\razor{} along the curve is five times that of \loremq{} (\qty{16}{} vs.\ \qty{3}{\milli\volt}). In contrast, our charge-conditioned long-range model \loremq{} follows the reference in all cases. Hence, charge-conditioned short-range models are limited by their effective cutoff, and field-conditioned ones by the homogeneous field near rough surfaces; long-range electrostatics lifts both limitations. For the long-range model, a scalar input suffices, because the model infers the surface normal from the atomic structure alone.

\prlsec{Charge-dependent properties} To test the models over a wider charge range, we perform \gls{md} of the Pt(111)--water interface with \grace{}-\razor{} and \loremq{}.
Both models are trained on data from Ref.~\citenum{bergmann_erasing_2026}, with energy, forces, work function, and surface-normal Born effective charges as targets.
A snapshot of the system is shown in Fig.~\ref{fig:capacitance}a. We compare with two references: our own \gls{aimd} (sampling time $2\times\qty{5}{ps}$ for each charge), and \grace{} models trained at a single charge each (\grace{} single-$q$).
For the latter reference, we generated additional training data by an active-learning procedure (see Sec.~\ref{sm:singleq} of the Supplemental Material for details).
Both references were obtained with the same computational settings as the literature data.
We leave out $q=+1\,e$ in training our \glspl{mlip}, as these data are prone to dielectric breakdown \citep{bergmann_erasing_2026}. As the work-function label, we use the work function of the solvated jellium method~\cite{kastlunger_controlled-potential_2018}, which differs from the charge derivative of the energy labels by about \qty{0.3}{\volt} (see Sec.~\ref{sm:labels} of the Supplemental Material for a comparison of both labels).

Both \grace{}-\razor{} and \loremq{} closely follow the water density and orientation of the references (Fig.~\ref{fig:capacitance}b--g). The agreement between the charge- and field-conditioned models is expected from the simple model system (Fig.~\ref{fig:pt}). The surface is flat, and all water in the \qty{10}{\angstrom} film has the electrode surface within the effective cutoff of both models (at most \qty{12}{\angstrom} for two-layer \grace{}). Both models also reproduce the \gls{aimd} electrode potential and the number of chemisorbed water molecules from $q=-1$ to $+0.75\,e$ (Fig.~\ref{fig:capacitance}h,i). % \todo{In the SI we discuss that part of this improvement is due to data-related decisions, not only model architecture. Nevertheless, our architectures enable accurate and fast simulations of charge-dependent properties.}

\prlsec{Conclusion} We have shown that a consistent \gls{mlip} of an electrified interface requires two ingredients: the bias as an input and long-range electrostatics.
Conditioning alone is not sufficient: across all test cases, from water between identical Ne electrodes to water above flat and stepped Pt, only the long-range model responds correctly and consistently.
At metal electrodes, a scalar electrode charge is a sufficient input for this model.
Field-conditioned short-range models, the current state of the art~\cite{bergmann_machine_2025,martin_general_2026}, remain limited by their effective cutoff.
The limitation also affects properties at the electrode surface: in constant-potential simulations, an incorrect electrode potential drives the potentiostat toward the wrong charge.
Our long-range model instead obtains the electrode potential from the same energy surface as the forces and can therefore be used directly in constant-potential simulations.
% Something like: Which counter charge best represents a real electrochemical cell remains open, since the charge there is screened by ions within a few \AA ngstr\"om of the interface and not by the jellium, plate or homogeneous background of our references. That choice sits in the reference calculations, not in the model: the bias enters as a single scalar, so any construction can be learned without touching the architecture.
Finally, our long-range models learn the electrostatic response without imposing its known functional form, Eq.~\eqref{eq:charging-integral}. Building such physical constraints into the architecture is a natural next step toward simulations of electrolytes with solvated ions and is expected to further improve the transferability to larger and more realistic systems.

% \begin{figure}[tb]
%   \centering
%   \includegraphics{figures/potential.pdf}
%   \caption{Displacement-field conditioning. The bias of a slab cell enters as $\mathbf{D}=\sigma\hat{\mathbf{n}}$, $\sigma=Q/A$. The magnitude modulates the invariant features $h_i$ of any backbone through \film; where symmetry requires it, the direction $\hat{\mathbf{n}}$ couples to the equivariant features $h_i^l$. Forces, work function, Born effective charges and capacitance follow by algorithmic differentiation of $U=\sum_i U_i$.
%   \todo{Redraw the $\hat{\mathbf{n}}$ path as conditional (needed only for symmetric cells).}}
%   \label{fig:model}
% \end{figure}

\prlsec{Acknowledgments}P.L. and P.R. acknowledge funding from the German Research Foundation (DFG) under Germany's Excellence Strategy -- EXC 2089/2 -- 390776260.
L.d.K. and C.J.S. acknowledge funding from the European Union; this publication is part of the HeliECat project that has received funding from the European Research Council (ERC) under the European Union's Horizon research and innovation program (Grant agreement No. 101219789). Views and opinions expressed are however those of the author(s) only and do not necessarily reflect those of the European Union or the European Research Council Executive Agency. Neither the European Union nor the granting authority can be held responsible for them.
The authors gratefully acknowledge the scientific support and HPC resources provided by the Leibniz Supercomputing Centre (LRZ) and by the Erlangen National High Performance Computing Center (NHR@FAU) of the Friedrich-Alexander-Universität Erlangen-Nürnberg (FAU) under NHR projects b311bb and b314bb.
The authors would like to thank Marcel F. Langer, Nicolas Bergmann, Henrik Stooß, Alexander Schlaich and Georg K. H. Madsen for insightful discussions.

\prlsec{Data availability}
The \lorem{} models have been trained using a branch of the public repository available at \url{https://github.com/lucasdekam/lorem-jax/tree/film-field-conditioning}. The code for training bias-conditioned \grace{} models is available at \url{https://github.com/lucasdekam/grace-tensorpotential/tree/charge-conditioning}. The modified version of \gls{lammps} with the \grace{} interface to enable constant-charge simulations is available at \url{https://github.com/lucasdekam/lammps/tree/grace}.
Further datasets, models, and analysis code to reproduce these findings will be made available upon publication in an open online data repository. % https://github.com/lucasdekam/charge-conditioning-long-range, create zenodo for code and one for the datasets.

\nocontentsline
\bibliography{philip, lucas}

@article{mathanker2026estimating,
  title={Estimating Potential-Dependent Physicochemical Properties at Metal--Electrolyte Interfaces Using Machine Learning Interatomic Potentials},
  author={Mathanker, Ankit and Guo, Jiawei and Goldsmith, Bryan R and Varley, Joel B and Govindarajan, Nitish},
  journal={ACS Electrochemistry},
  volume={2},
  number={5},
  pages={1176--1189},
  year={2026},
  publisher={ACS Publications}
}

@article{zhang2026electrochemical,
  title={Electrochemical electron transfer: Key concepts, theories, and parameterization via atomistic simulations},
  author={Zhang, Mengke and Chen, Yanxia and Melander, Marko M and Huang, Jun},
  journal={Chemical Reviews},
  volume={126},
  number={13},
  pages={7407},
  year={2026}
}

@article{sechi2025first,
  title={First-Principles Molecular Dynamics with Potential and Charge Fluctuations Applied to Au (111) in Alkaline Solutions},
  author={Sechi, Renata and Kastlunger, Georg and Bhowmik, Arghya and Hansen, Heine Anton},
  journal={Journal of Chemical Theory and Computation},
  volume={21},
  number={10},
  pages={5279--5290},
  year={2025},
  publisher={ACS Publications}
}

@article{bonnet2012first,
  title={First-principles molecular dynamics at a constant electrode potential},
  author={Bonnet, Nic{\'e}phore and Morishita, Tetsuya and Sugino, Osamu and Otani, Minoru},
  journal={Physical Review Letters},
  volume={109},
  number={26},
  pages={266101},
  year={2012},
  publisher={APS}
}

@article{hjorth2017atomic,
  title={The atomic simulation environment—a Python library for working with atoms},
  author={Hjorth Larsen, Ask and J{\o}rgen Mortensen, Jens and Blomqvist, Jakob and Castelli, Ivano E and Christensen, Rune and Du{\l}ak, Marcin and Friis, Jesper and Groves, Michael N and Hammer, Bj{\o}rk and Hargus, Cory and others},
  journal={Journal of Physics: Condensed Matter},
  volume={29},
  number={27},
  pages={273002},
  year={2017},
  publisher={IOP Publishing}
}

@article{bussi2007canonical,
  title={Canonical sampling through velocity rescaling},
  author={Bussi, Giovanni and Donadio, Davide and Parrinello, Michele},
  journal={The Journal of chemical physics},
  volume={126},
  number={1},
  year={2007},
  publisher={AIP Publishing}
}

@article{mortensen2024gpaw,
  title={GPAW: An open Python package for electronic structure calculations},
  author={Mortensen, Jens J{\o}rgen and Larsen, Ask Hjorth and Kuisma, Mikael and Ivanov, Aleksei V and Taghizadeh, Alireza and Peterson, Andrew and Haldar, Anubhab and Dohn, Asmus Ougaard and Sch{\"a}fer, Christian and J{\'o}nsson, Elvar {\"O}rn and others},
  journal={The Journal of Chemical Physics},
  volume={160},
  number={9},
  year={2024},
  publisher={AIP Publishing}
}

@article{yu2026maximizing,
  title={Maximizing efficiency of dataset compression for machine learning potentials with information theory},
  author={Yu, Benjamin and Lordi, Vincenzo and Schwalbe-Koda, Daniel},
  journal={The Journal of Chemical Physics},
  volume={164},
  number={16},
  year={2026},
  publisher={AIP Publishing}
}

@article{surendralal2018first,
  title={First-principles approach to model electrochemical reactions: understanding the fundamental mechanisms behind Mg corrosion},
  author={Surendralal, Sudarsan and Todorova, Mira and Finnis, Michael W and Neugebauer, J{\"o}rg},
  journal={Physical Review Letters},
  volume={120},
  number={24},
  pages={246801},
  year={2018},
  publisher={APS}
}

@article{kresse1993ab,
  title={Ab initio molecular dynamics for liquid metals},
  author={Kresse, Georg and Hafner, J{\"u}rgen},
  journal={Physical review B},
  volume={47},
  number={1},
  pages={558},
  year={1993},
  publisher={APS},
  doi={10.1016/0022-3093(95)00355-X},
}

@article{kresse1994ab,
  title={Ab initio molecular-dynamics simulation of the liquid-metal--amorphous-semiconductor transition in germanium},
  author={Kresse, Georg and Hafner, J{\"u}rgen},
  journal={Physical Review B},
  volume={49},
  number={20},
  pages={14251},
  year={1994},
  publisher={APS},
  doi={10.1103/PhysRevB.49.14251},
}

@article{kresse1996efficiency,
  title={Efficiency of ab-initio total energy calculations for metals and semiconductors using a plane-wave basis set},
  author={Kresse, Georg and Furthm{\"u}ller, J{\"u}rgen},
  journal={Computational materials science},
  volume={6},
  number={1},
  pages={15--50},
  year={1996},
  publisher={Elsevier},
  doi={10.1016/0927-0256(96)00008-0},
}

@article{kresse1996efficient,
  title={Efficient iterative schemes for ab initio total-energy calculations using a plane-wave basis set},
  author={Kresse, Georg and Furthm{\"u}ller, J{\"u}rgen},
  journal={Physical review B},
  volume={54},
  number={16},
  pages={11169},
  year={1996},
  publisher={APS},
  doi={10.1103/PhysRevB.54.11169},
}

@article{kresse1999ultrasoft,
  title={From ultrasoft pseudopotentials to the projector augmented-wave method},
  author={Kresse, Georg and Joubert, Daniel},
  journal={Physical review b},
  volume={59},
  number={3},
  pages={1758},
  year={1999},
  publisher={APS},
  doi={10.1103/PhysRevB.59.1758},
}

@article{grimme2010consistent,
    author = {Grimme, Stefan and Antony, Jens and Ehrlich, Stephan and Krieg, Helge},
    journal = {The Journal of chemical physics},
    number = {15},
    publisher = {AIP Publishing},
    title = {A consistent and accurate ab initio parametrization of density functional dispersion correction {(DFT-D)} for the 94 elements {H-Pu}},
    volume = {132},
    year = {2010}
}

@article{hammer1999improved,
    author = {Hammer, BHLB and Hansen, Lars Bruno and N{\o}rskov, Jens Kehlet},
    doi = {10.1103/PhysRevB.59.7413},
    journal = {Physical review B},
    number = {11},
    pages = {7413},
    publisher = {APS},
    title = {Improved adsorption energetics within density-functional theory using revised {Perdew-Burke-Ernzerhof} functionals},
    volume = {59},
    year = {1999}
}

@article{monkhorst1976special,
    author = {Monkhorst, Hendrik J and Pack, James D},
    doi = {10.1103/PhysRevB.16.1748},
    journal = {Physical review B},
    number = {12},
    pages = {5188},
    publisher = {APS},
    title = {Special points for {Brillouin}-zone integrations},
    volume = {13},
    year = {1976}
}

@article{perdew1996generalized,
  title={Generalized gradient approximation made simple},
  author={Perdew, John P and Burke, Kieron and Ernzerhof, Matthias},
  journal={Physical review letters},
  volume={77},
  number={18},
  pages={3865},
  year={1996},
  publisher={APS}
}

@article{schlipf2015optimization,
  title={Optimization algorithm for the generation of ONCV pseudopotentials},
  author={Schlipf, Martin and Gygi, Fran{\c{c}}ois},
  journal={Computer Physics Communications},
  volume={196},
  pages={36--44},
  year={2015},
  publisher={Elsevier}
}

@article{hamann2013optimized,
  title={Optimized norm-conserving Vanderbilt pseudopotentials},
  author={Hamann, Donald R},
  journal={Physical Review B—Condensed Matter and Materials Physics},
  volume={88},
  number={8},
  pages={085117},
  year={2013},
  publisher={APS}
}

@article{schwalbekoda2025information,
    title = {Model-free estimation of completeness, uncertainties, and outliers in atomistic machine learning using information theory},
    author = {Schwalbe-Koda, Daniel and Hamel, Sebastien and Sadigh, Babak and Zhou, Fei and Lordi, Vincenzo},
    year = {2025},
    journal = {Nature Communications},
    url = {https://doi.org/10.1038/s41467-025-59232-0},
    doi = {10.1038/s41467-025-59232-0},
    volume={16},
    pages={4014},
}

@article{bartok2013representing,
  title={On representing chemical environments},
  author={Bart{\'o}k, Albert P and Kondor, Risi and Cs{\'a}nyi, G{\'a}bor},
  journal={Physical Review B—Condensed Matter and Materials Physics},
  volume={87},
  number={18},
  pages={184115},
  year={2013},
  publisher={APS}
}

@article{bigi_metatensor_2026,
  title = {Metatensor and Metatomic: {{Foundational}} Libraries for Interoperable Atomistic Machine Learning},
  shorttitle = {Metatensor and Metatomic},
  author = {Bigi, Filippo and Abbott, Joseph W. and Loche, Philip and Mazitov, Arslan and Tisi, Davide and Langer, Marcel F. and Goscinski, Alexander and Pegolo, Paolo and Chong, Sanggyu and Goswami, Rohit and Febrer, Pol and Chorna, Sofiia and Kellner, Matthias and Ceriotti, Michele and Fraux, Guillaume},
  year = 2026,
  month = feb,
  journal = {J. Chem. Phys.},
  volume = {164},
  number = {6},
  pages = {064113},
  issn = {0021-9606},
  doi = {10.1063/5.0304911},
}

@article{aouichaoui_machine_2026,
  title = {Machine Learning Potentials for Electrified Interfaces},
  author = {Aouichaoui, Adem R. N. and Tisi, Davide and Castelli, Ivano E. and Bhowmik, Arghya},
  year = 2026,
  month = jun,
  journal = {RSC Appl. Interfaces},
  issn = {2755-3701},
  doi = {10.1039/d6lf00149a},
  urldate = {2026-07-28}
}

@misc{batatia_mace-polar-1_2026,
  title = {{{MACE-POLAR-1}}: {{A Polarisable Electrostatic Foundation Model}} for {{Molecular Chemistry}}},
  shorttitle = {{{MACE-POLAR-1}}},
  author = {Batatia, Ilyes and Baldwin, William J. and Kuryla, Domantas and Hart, Joseph and Kasoar, Elliott and Elena, Alin M. and Moore, Harry and Gawkowski, Miko{\l}aj J. and Shi, Benjamin X. and Kapil, Venkat and Kourtis, Panagiotis and Magd{\u a}u, Ioan-Bogdan and Cs{\'a}nyi, G{\'a}bor},
  year = 2026,
  month = feb,
  number = {arXiv:2602.19411},
  eprint = {2602.19411},
  primaryclass = {physics},
  publisher = {arXiv},
  doi = {10.48550/arXiv.2602.19411},
  urldate = {2026-04-01},
  archiveprefix = {arXiv}
}

@article{bergmann_erasing_2026,
  title = {Erasing Dielectric Breakdown Artifacts to Machine-Learn Charged {{Pt}}--Water Interfaces},
  author = {Bergmann, Nicolas and Reuter, Karsten and H{\"o}rmann, Nicolas G.},
  year = 2026,
  month = may,
  journal = {J. Chem. Phys.},
  volume = {164},
  number = {17},
  pages = {174110},
  issn = {0021-9606, 1089-7690},
  doi = {10.1063/5.0323247},
  urldate = {2026-11-10},
  langid = {english}
}

@article{bergmann_machine_2025,
  title = {Machine {{Learning}} the {{Energetics}} of {{Electrified Solid-Liquid Interfaces}}},
  author = {Bergmann, Nicolas and Bonnet, Nic{\'e}phore and Marzari, Nicola and Reuter, Karsten and H{\"o}rmann, Nicolas G.},
  year = 2025,
  month = sep,
  journal = {Phys. Rev. Lett.},
  volume = {135},
  number = {14},
  pages = {146201},
  publisher = {American Physical Society},
  doi = {10.1103/lm64-m3bn},
  urldate = {2026-01-06}
}

@article{bochkarev_graph_2024,
  title = {Graph {{Atomic Cluster Expansion}} for {{Semilocal Interactions}} beyond {{Equivariant Message Passing}}},
  author = {Bochkarev, Anton and Lysogorskiy, Yury and Drautz, Ralf},
  year = 2024,
  month = jun,
  journal = {Phys. Rev. X},
  volume = {14},
  number = {2},
  pages = {021036},
  issn = {2160-3308},
  doi = {10.1103/PhysRevX.14.021036},
  urldate = {2026-11-10},
  langid = {english}
}

@misc{brezina_electrochemical_2026,
  title = {Electrochemical {{Molecular Dynamics}} within the {{Finite-Field Framework}}},
  author = {Brezina, Krystof and {Ahrens-Iwers}, Ludwig and Rossi, Mariana},
  year = 2026,
  month = oct,
  number = {arXiv:2610.23291},
  eprint = {2610.23291},
  primaryclass = {physics.chem-ph},
  publisher = {arXiv},
  doi = {10.48550/arXiv.2610.23291},
  urldate = {2026-10-28},
  archiveprefix = {arXiv}
}

@article{chen_constant-potential_2026,
  title = {Constant-{{Potential MD}} with {{Neural Network Potentials Reveals Cation Effects}} on {{CO}}{\textsubscript{2}} {{Reduction}} at {{Au-Water Interfaces}}},
  author = {Chen, Letian and Tian, Yun and Hu, Xu and Chen, Suya and Wang, Huijuan and Zhang, Xu and Zhou, Zhen},
  year = 2026,
  month = feb,
  journal = {JACS Au},
  volume = {6},
  number = {2},
  pages = {868--880},
  issn = {2691-3704, 2691-3704},
  doi = {10.1021/jacsau.5c01198},
  urldate = {2026-11-10},
  langid = {english}
}

@article{de_kam_modeling_2026,
  title = {Modeling the Inner Electric Double Layer Using Short-Range Machine Learning Potentials: {{A}} Critical Assessment},
  shorttitle = {Modeling the Inner Electric Double Layer Using Short-Range Machine Learning Potentials},
  author = {{de Kam}, Lucas B. T. and Zhu, Jia-Xin and Mathanker, Ankit and {Doblhoff-Dier}, Katharina and Govindarajan, Nitish},
  year = 2026,
  month = nov,
  journal = {J. Chem. Phys.},
  volume = {165},
  number = {18},
  pages = {184709},
  issn = {0021-9606},
  doi = {10.1063/5.0359674},
  urldate = {2026-11-11}
}

@article{deisenbeck_dielectric_2023,
  title = {Dielectric {{Properties}} of {{Nanoconfined Water}} from {{Ab Initio Thermopotentiostat Molecular Dynamics}}},
  author = {Dei{\ss}enbeck, Florian and Wippermann, Stefan},
  year = 2023,
  month = feb,
  journal = {J. Chem. Theory Comput.},
  volume = {19},
  number = {3},
  pages = {1035--1043},
  publisher = {American Chemical Society},
  issn = {1549-9618},
  doi = {10.1021/acs.jctc.2c00959},
  urldate = {2026-03-17}
}

@article{falletta_unified_2025,
  title = {Unified Differentiable Learning of Electric Response},
  author = {Falletta, Stefano and Cepellotti, Andrea and Johansson, Anders and Tan, Chuin Wei and Descoteaux, Marc L. and Musaelian, Albert and Owen, Cameron J. and Kozinsky, Boris},
  year = 2025,
  month = apr,
  journal = {Nat. Commun.},
  volume = {16},
  number = {1},
  pages = {4031},
  publisher = {Nature Publishing Group},
  issn = {2041-1723},
  doi = {10.1038/s41467-025-59304-1},
  urldate = {2026-04-16},
  langid = {english}
}

@article{gao_self-consistent_2022,
  title = {Self-Consistent Determination of Long-Range Electrostatics in Neural Network Potentials},
  author = {Gao, Ang and Remsing, Richard C.},
  year = 2022,
  month = mar,
  journal = {Nat. Commun.},
  volume = {13},
  number = {1},
  pages = {1572},
  publisher = {Nature Publishing Group},
  issn = {2041-1723},
  doi = {10.1038/s41467-022-29243-2},
  urldate = {2022-12-19},
  langid = {english}
}

@misc{grasselli_long-range_2026,
  title = {Long-Range Electrostatics in Atomistic Machine Learning: A Physical Perspective},
  shorttitle = {Long-Range Electrostatics in Atomistic Machine Learning},
  author = {Grasselli, Federico and Rossi, Kevin and de Gironcoli, Stefano and Grisafi, Andrea},
  year = 2026,
  month = feb,
  number = {arXiv:2602.11071},
  eprint = {2602.11071},
  primaryclass = {cond-mat.mtrl-sci},
  publisher = {arXiv},
  doi = {10.48550/arXiv.2602.11071},
  urldate = {2026-07-15},
  archiveprefix = {arXiv}
}

@article{grimme_effect_2011,
  title = {Effect of the Damping Function in Dispersion Corrected Density Functional Theory},
  author = {Grimme, Stefan and Ehrlich, Stephan and Goerigk, Lars},
  year = 2011,
  journal = {J. Comput. Chem.},
  volume = {32},
  number = {7},
  pages = {1456--1465},
  issn = {1096-987X},
  doi = {10.1002/jcc.21759},
  urldate = {2026-11-15},
  langid = {english}
}

@article{grisafi_predicting_2023,
  title = {Predicting the Charge Density Response in Metal Electrodes},
  author = {Grisafi, Andrea and Bussy, Augustin and Salanne, Mathieu and Vuilleumier, Rodolphe},
  year = 2023,
  month = dec,
  journal = {Phys. Rev. Mater.},
  volume = {7},
  number = {12},
  pages = {125403},
  publisher = {American Physical Society},
  doi = {10.1103/PhysRevMaterials.7.125403},
  urldate = {2024-04-30}
}

@article{hagopian_assessing_2026,
  title = {Assessing the Potential of Zero Charge in Ab Initio Molecular Dynamics Simulations},
  author = {Hagopian, Arthur and Koper, Marc T. M. and {Doblhoff-Dier}, Katharina},
  year = 2026,
  month = may,
  journal = {J. Chem. Phys.},
  volume = {164},
  number = {17},
  pages = {174711},
  issn = {0021-9606, 1089-7690},
  doi = {10.1063/5.0322322},
  urldate = {2026-11-10},
  langid = {english}
}

@article{joll_machine_2024,
  title = {Machine Learning the Electric Field Response of Condensed Phase Systems Using Perturbed Neural Network Potentials},
  author = {Joll, Kit and Schienbein, Philipp and Rosso, Kevin M. and Blumberger, Jochen},
  year = 2024,
  month = sep,
  journal = {Nat. Commun.},
  volume = {15},
  number = {1},
  pages = {8192},
  publisher = {Nature Publishing Group},
  issn = {2041-1723},
  doi = {10.1038/s41467-024-52491-3},
  urldate = {2025-03-21},
  langid = {english}
}

@article{kastlunger_controlled-potential_2018,
  title = {Controlled-{{Potential Simulation}} of {{Elementary Electrochemical Reactions}}: {{Proton Discharge}} on {{Metal Surfaces}}},
  shorttitle = {Controlled-{{Potential Simulation}} of {{Elementary Electrochemical Reactions}}},
  author = {Kastlunger, Georg and Lindgren, Per and Peterson, Andrew A.},
  year = 2018,
  month = jun,
  journal = {J. Phys. Chem. C},
  volume = {122},
  number = {24},
  pages = {12771--12781},
  issn = {1932-7447, 1932-7455},
  doi = {10.1021/acs.jpcc.8b02465},
  urldate = {2026-11-10},
  langid = {english}
}

@article{ko_fourth-generation_2021,
  title = {A Fourth-Generation High-Dimensional Neural Network Potential with Accurate Electrostatics Including Non-Local Charge Transfer},
  author = {Ko, Tsz Wai and Finkler, Jonas A. and Goedecker, Stefan and Behler, J{\"o}rg},
  year = 2021,
  month = jan,
  journal = {Nat. Commun.},
  volume = {12},
  number = {1},
  pages = {398},
  publisher = {Nature Publishing Group},
  issn = {2041-1723},
  doi = {10.1038/s41467-020-20427-2},
  urldate = {2021-12-08},
  langid = {english}
}

@article{li_electron_2025,
  title = {Electron {{Spillover}} into {{Water Layers}}: {{A Quantum Leap}} in {{Understanding Capacitance Behavior}}},
  shorttitle = {Electron {{Spillover}} into {{Water Layers}}},
  author = {Li, Lang and Eggert, Thorben and Reuter, Karsten and H{\"o}rmann, Nicolas G.},
  year = 2025,
  month = jul,
  journal = {J. Am. Chem. Soc.},
  volume = {147},
  number = {26},
  pages = {22778--22784},
  issn = {0002-7863, 1520-5126},
  doi = {10.1021/jacs.5c04728},
  urldate = {2026-11-10},
  langid = {english}
}

@misc{liu_muon_2025,
  title = {Muon Is {{Scalable}} for {{LLM Training}}},
  author = {Liu, Jingyuan and Su, Jianlin and Yao, Xingcheng and Jiang, Zhejun and Lai, Guokun and Du, Yulun and Qin, Yidao and Xu, Weixin and Lu, Enzhe and Yan, Junjie and Chen, Yanru and Zheng, Huabin and Liu, Yibo and Liu, Shaowei and Yin, Bohong and He, Weiran and Zhu, Han and Wang, Yuzhi and Wang, Jianzhou and Dong, Mengnan and Zhang, Zheng and Kang, Yongsheng and Zhang, Hao and Xu, Xinran and Zhang, Yutao and Wu, Yuxin and Zhou, Xinyu and Yang, Zhilin},
  year = 2025,
  publisher = {arXiv},
  doi = {10.48550/ARXIV.2502.16982},
  urldate = {2026-11-15}
}

@article{martin_general_2026,
  title = {General {{Learning}} of the {{Electric Response}} of {{Inorganic Materials}}},
  author = {Martin, Bradley A. A. and Ganose, Alex M. and Kapil, Venkat and Li, Tingwei and Butler, Keith T.},
  year = 2026,
  month = jul,
  journal = {PRX Intell.},
  volume = {1},
  number = {1},
  pages = {013006},
  publisher = {American Physical Society},
  doi = {10.1103/b116-xy8k},
  urldate = {2026-08-04}
}

@article{mattoso_user-defined_2026,
  title = {User-{{Defined Electrostatic Potentials}} in {{DFT Supercell Calculations}}: {{Implementation}} and {{Application}} to {{Electrified Interfaces}}},
  shorttitle = {User-{{Defined Electrostatic Potentials}} in {{DFT Supercell Calculations}}},
  author = {Mattoso, Samuel and Yang, Jing and Dei{\ss}enbeck, Florian and Abdelkawy, Ahmed and Freysoldt, Christoph and Wippermann, Stefan and Todorova, Mira and Neugebauer, J{\"o}rg},
  year = 2026,
  month = mar,
  journal = {J. Chem. Theory Comput.},
  volume = {22},
  number = {5},
  pages = {2597--2606},
  issn = {1549-9618, 1549-9626},
  doi = {10.1021/acs.jctc.5c02150},
  urldate = {2026-11-10},
  langid = {english}
}

@misc{parker_false_2026,
  title = {False {{Metallization}} in {{Short-Ranged Machine Learned Interatomic Potentials}}},
  author = {Parker, Isaac J. and Hoffmann, Mandy J. and Baldwin, William J. and Han, Shuang and Gupta, Srishti and Fong, Kara D. and Michaelides, Angelos and Schran, Christoph and De, Sandip and Cs{\'a}nyi, G{\'a}bor},
  year = 2026,
  month = mar,
  number = {arXiv:2603.04228},
  eprint = {2603.04228},
  primaryclass = {physics.chem-ph},
  publisher = {arXiv},
  doi = {10.48550/arXiv.2603.04228},
  urldate = {2026-06-26},
  archiveprefix = {arXiv}
}

@article{perez_film_2018,
  title = {{{FiLM}}: {{Visual Reasoning}} with a {{General Conditioning Layer}}},
  shorttitle = {{{FiLM}}},
  author = {Perez, Ethan and Strub, Florian and De Vries, Harm and Dumoulin, Vincent and Courville, Aaron},
  year = 2018,
  month = apr,
  journal = {Proc. AAAI Conf. Artif. Intell.},
  volume = {32},
  number = {1},
  issn = {2374-3468, 2159-5399},
  doi = {10.1609/aaai.v32i1.11671},
  urldate = {2026-11-10}
}

@article{rumiantsev_learning_2026,
  title = {Learning Long-Range Representations with Equivariant Messages},
  author = {Rumiantsev, Egor and Langer, Marcel F. and Sodjargal, Tulga-Erdene and Ceriotti, Michele and Loche, Philip},
  year = 2026,
  journal = {Trans. Mach. Learn. Res.},
  issn = {2835-8856}
}

@misc{sahel-schackis_equifilm_2026,
  title = {{{EquiFiLM}}: {{Charge-Conditioned Equivariant Force Fields}} via {{Feature}}-Wise {{Linear Modulation}}},
  shorttitle = {{{EquiFiLM}}},
  author = {{Sahel-Schackis}, Samuel and Nomura, Ken-ichi and Nakano, Aiichiro and Kling, Matthias F. and Linker, Thomas},
  year = 2026,
  month = jul,
  number = {arXiv:2607.05559},
  eprint = {2607.05559},
  publisher = {arXiv},
  doi = {10.48550/ARXIV.2607.05559},
  urldate = {2026-11-10},
  archiveprefix = {arXiv}
}

@article{siepmann_influence_1995,
  title = {Influence of Surface Topology and Electrostatic Potential on Water/Electrode Systems},
  author = {Siepmann, J. Ilja and Sprik, Michiel},
  year = 1995,
  month = jan,
  journal = {J. Chem. Phys.},
  volume = {102},
  number = {1},
  pages = {511--524},
  publisher = {American Institute of Physics},
  issn = {0021-9606},
  doi = {10.1063/1.469429},
  urldate = {2020-07-29},
  langid = {english}
}

@article{stengel_electric_2009,
  title = {Electric Displacement as the Fundamental Variable in Electronic-Structure Calculations},
  author = {Stengel, Massimiliano and Spaldin, Nicola A. and Vanderbilt, David},
  year = 2009,
  month = apr,
  journal = {Nat. Phys.},
  volume = {5},
  number = {4},
  pages = {304--308},
  publisher = {Nature Publishing Group},
  issn = {1745-2481},
  doi = {10.1038/nphys1185},
  urldate = {2026-07-14},
  langid = {english}
}

@misc{sumic_electrostatic_2026,
  title = {Electrostatic {{Phenomenology Benchmarks}} for {{Machine-Learned Interatomic Potentials}} in {{Electrochemistry}}: {{Beyond}} the {{Energy-Force Metric}}},
  shorttitle = {Electrostatic {{Phenomenology Benchmarks}} for {{Machine-Learned Interatomic Potentials}} in {{Electrochemistry}}},
  author = {Sumi{\'c}, Barbara and Vasdev, Ria and Ethirajan, Sudheesh Kumar and Yang, Jing and Cucinotta, Clotilde S. and Hennig, Richard G. and Reuter, Karsten and Ringe, Stefan and Todorova, Mira and Freysoldt, Christoph and Neugebauer, J{\"o}rg},
  year = 2026,
  month = aug,
  number = {arXiv:2608.14153},
  eprint = {2608.14153},
  primaryclass = {cond-mat.mtrl-sci},
  publisher = {arXiv},
  doi = {10.48550/arXiv.2608.14153},
  urldate = {2026-08-17},
  archiveprefix = {arXiv}
}

@article{sun_electrochemical_2026,
  title = {Electrochemical {{Potential Fluctuation Matters}} in {{Rate Constant Calculations}} for {{Proton-Coupled Electron Transfer}}},
  author = {Sun, Menglin and Fu, Li and Xu, Shenzhen},
  year = 2026,
  month = feb,
  journal = {J. Chem. Theory Comput.},
  volume = {22},
  number = {4},
  pages = {1539--1548},
  issn = {1549-9618},
  doi = {10.1021/acs.jctc.5c01534},
  urldate = {2026-11-03}
}

@article{sun_probing_2025,
  title = {Probing Nuclear Quantum Effects in Electrocatalysis via a Machine-Learning Enhanced Grand Canonical Constant Potential Approach},
  author = {Sun, Menglin and Jin, Bin and Yang, Xiaolong and Xu, Shenzhen},
  year = 2025,
  month = apr,
  journal = {Nat. Commun.},
  volume = {16},
  number = {1},
  pages = {3600},
  issn = {2041-1723},
  doi = {10.1038/s41467-025-58871-7},
  urldate = {2026-11-10},
  langid = {english}
}

@article{unke_spookynet_2021,
  title = {{{SpookyNet}}: {{Learning}} Force Fields with Electronic Degrees of Freedom and Nonlocal Effects},
  shorttitle = {{{SpookyNet}}},
  author = {Unke, Oliver T. and Chmiela, Stefan and Gastegger, Michael and Sch{\"u}tt, Kristof T. and Sauceda, Huziel E. and M{\"u}ller, Klaus-Robert},
  year = 2021,
  month = dec,
  journal = {Nat. Commun.},
  volume = {12},
  number = {1},
  pages = {7273},
  issn = {2041-1723},
  doi = {10.1038/s41467-021-27504-0},
  urldate = {2026-11-10},
  langid = {english}
}

@article{wang_constant-potential_2025,
  title = {Constant-{{Potential Machine Learning Force Field}} for the {{Electrochemical Interface}}},
  author = {Wang, Ruoyu and Fang, Shaoheng and Huang, Qixing and Liu, Yuanyue},
  year = 2025,
  month = jul,
  journal = {J. Chem. Theory Comput.},
  volume = {21},
  number = {15},
  pages = {7628--7635},
  issn = {1549-9618},
  doi = {10.1021/acs.jctc.5c00784},
  urldate = {2026-09-09}
}

@article{wang_ion-modulated_2026,
  title = {Ion-Modulated Structure, Proton Transfer, and Capacitance in the {{Pt}}-Water Electric Double Layer},
  author = {Wang, Xiaoyu and Chen, Junmin and Zeng, Zezhu and Stein, Frederick and Lim, Junho and Cheng, Bingqing},
  year = 2026,
  month = jul,
  journal = {Npj Comput. Mater.},
  publisher = {Nature Publishing Group},
  issn = {2057-3960},
  doi = {10.1038/s41524-026-02230-7},
  urldate = {2026-08-06},
  langid = {english}
}

@article{xie_incorporating_2020,
  title = {Incorporating {{Electronic Information}} into {{Machine Learning Potential Energy Surfaces}} via {{Approaching}} the {{Ground-State Electronic Energy}} as a {{Function}} of {{Atom-Based Electronic Populations}}},
  author = {Xie, Xiaowei and Persson, Kristin A. and Small, David W.},
  year = 2020,
  month = jul,
  journal = {J. Chem. Theory Comput.},
  volume = {16},
  number = {7},
  pages = {4256--4270},
  publisher = {American Chemical Society},
  issn = {1549-9618},
  doi = {10.1021/acs.jctc.0c00217},
  urldate = {2022-12-19},
  langid = {english}
}

@article{yang_macro-dipole-constrained_2026,
  title = {Macro-{{Dipole-Constrained Learning}} of {{Atomic Charges}} for {{Accurate Electrostatic Potentials}} at {{Electrochemical Interfaces}}},
  author = {Yang, Jing and Li, Bingxin and Mattoso, Samuel and Abdelkawy, Ahmed and Todorova, Mira and Neugebauer, J{\"o}rg},
  year = 2026,
  month = mar,
  journal = {Phys. Rev. Lett.},
  volume = {136},
  number = {11},
  pages = {118001},
  publisher = {American Physical Society},
  doi = {10.1103/5hh2-wypm},
  urldate = {2026-10-20}
}

@article{yeh_ewald_1999,
  title = {Ewald Summation for Systems with Slab Geometry},
  author = {Yeh, In-Chul and Berkowitz, Max L.},
  year = 1999,
  journal = {J. Chem. Phys.},
  volume = {111},
  number = {7},
  pages = {3155--3162},
  doi = {10.1063/1.479595},
  langid = {english}
}

@article{zhang_finite_2016,
  title = {Finite Field Methods for the Supercell Modeling of Charged Insulator/Electrolyte Interfaces},
  author = {Zhang, Chao and Sprik, Michiel},
  year = 2016,
  month = dec,
  journal = {Phys. Rev. B},
  volume = {94},
  number = {24},
  pages = {245309},
  publisher = {American Physical Society},
  doi = {10.1103/PhysRevB.94.245309},
  urldate = {2026-03-17}
}

@article{zhou_abacus_2025,
  title = {{{ABACUS}}: {{An}} Electronic Structure Analysis Package for the {{AI}} Era},
  shorttitle = {{{ABACUS}}},
  author = {Zhou, Weiqing and Zheng, Daye and Liu, Qianrui and Lu, Denghui and Liu, Yu and Lin, Peize and Huang, Yike and Peng, Xingliang and Bao, Jie J. and Cai, Chun and Jin, Zuxin and Wu, Jing and Zhang, Haochong and Jin, Gan and Ji, Yuyang and Shen, Zhenxiong and Liu, Xiaohui and Sun, Liang and Cao, Yu and Sun, Menglin and Liu, Jianchuan and Chen, Tao and Liu, Renxi and Li, Yuanbo and Han, Haozhi and Liang, Xinyuan and Bao, Taoni and Deng, Zichao and Liu, Tao and Chen, Nuo and Ren, Hongxu and Zhang, Xiaoyang and Liu, Zhaoqing and Fu, Yiwei and Liu, Maochang and Li, Zhuoyuan and Wen, Tongqi and Tang, Zechen and Xu, Yong and Duan, Wenhui and Wang, Xiaoyang and Gu, Qiangqiang and Dai, Fu-Zhi and Zheng, Qijing and Zhong, Yang and Xiang, Hongjun and Gong, Xingao and Zhao, Jin and Zhang, Yuzhi and Ou, Qi and Jiang, Hong and Liu, Shi and Xu, Ben and Xu, Shenzhen and Ren, Xinguo and He, Lixin and Zhang, Linfeng and Chen, Mohan},
  year = 2025,
  month = nov,
  journal = {J. Chem. Phys.},
  volume = {163},
  number = {19},
  pages = {192501},
  issn = {0021-9606, 1089-7690},
  doi = {10.1063/5.0297563},
  urldate = {2026-11-10},
  langid = {english}
}

@article{zhu_machine_2025,
  title = {Machine {{Learning Potential}} for {{Electrochemical Interfaces}} with {{Hybrid Representation}} of {{Dielectric Response}}},
  author = {Zhu, Jia-Xin and Cheng, Jun},
  year = 2025,
  month = jul,
  journal = {Phys. Rev. Lett.},
  volume = {135},
  number = {1},
  pages = {018003},
  publisher = {American Physical Society},
  doi = {10.1103/48ct-3jxm},
  urldate = {2026-10-21}
}
\restorecontentsline

% ===========================================================================
% Supplemental Material (kept here while drafting; split into a separate
% file for submission)
% ===========================================================================
\clearpage
\onecolumngrid
\setcounter{page}{1}
\renewcommand{\thepage}{S\arabic{page}}
\begin{center}
  {\large\bfseries Supplemental Material:\\[0.3em]
  Bias Conditioning and Long-Range Electrostatics for Consistent Machine-Learning Potentials of Electrified Interfaces\par}
  \vspace{1em}
  Lucas B. T. de Kam,$^{1,2}$ Patrick Rinke,$^{2,3,4}$ Christopher J. Stein,$^{1,2}$ and Philip Loche$^{2,3,4,*}$\par
  \vspace{0.8em}
  {\small\itshape
  $^{1}$Department of Chemistry, School of Natural Sciences, Technical University of Munich, Garching, Germany\\
  $^{2}$Atomistic Modeling Center, Munich Data Science Institute, Technical University of Munich, Garching, Germany\\
  $^{3}$Department of Physics, School of Natural Sciences, Technical University of Munich, Garching, Germany\\
  $^{4}$Munich Center for Machine Learning (MCML)\par}
  \vspace{0.5em}
  {\small E-mail: \href{mailto:lucas.de.kam@tum.de}{lucas.de.kam@tum.de},
  \href{mailto:patrick.rinke@tum.de}{patrick.rinke@tum.de},
  \href{mailto:christopher.stein@tum.de}{christopher.stein@tum.de},
  \href{mailto:philip.loche@tum.de}{philip.loche@tum.de}\par
  $^{*}$Corresponding author\par}
\end{center}
\vspace{1em}
\tableofcontents
\clearpage
\setcounter{equation}{0}
\setcounter{figure}{0}
\setcounter{table}{0}
\setcounter{section}{0}
\renewcommand{\theequation}{S\arabic{equation}}
\renewcommand{\thefigure}{S\arabic{figure}}
\renewcommand{\thetable}{S\arabic{table}}
\renewcommand{\thesection}{S\arabic{section}}
\makeatletter\renewcommand{\p@subsection}{\thesection}\makeatother % refs read ``S1A''
\setcounter{secnumdepth}{2}
\glsresetall % the Supplemental Material defines acronyms anew

\section{Water between Ne electrodes} \label{sm:ne-capacitor}

\subsection{Energy of the Ne--water--Ne capacitor under applied field} \label{sm:newaterne-energy}
We label the Ne--water--Ne cell under an external field $E_0\hat{\mathbf{z}}$ with a slab dipole correction. The cell is periodic in $x$ and $y$, has length $L$ along $z$ and a volume $\Omega=AL$. The displacement field is $\mathbf{D}=\varepsilon_0 E_0 \hat{\mathbf{z}}$ throughout the cell. At fixed $\mathbf{D}$, the energy of a periodic, insulating cell is~\citep{stengel_electric_2009}
\begin{equation}\label{eq:stengel}
    U(\mathbf{D}) = U_\mathrm{PBC} + \frac{\Omega}{2\varepsilon_0}|\mathbf{D} - \mathbf{P}|^2,
\end{equation}
where $U_\mathrm{PBC}$ is the energy under ordinary periodic boundary conditions. The polarization $\mathbf{P}$ is related to the dipole moment $\mathbf{M}$ as $\mathbf{M}=\Omega\mathbf{P}$. With $\mathbf{D}$ and $\mathbf{P}$ along $z$, expanding the square gives
\begin{equation} \label{eq:sm-expanded}
    U = \underbrace{U_\mathrm{PBC} + \frac{M_z^2}{2\varepsilon_0\Omega}}_{U_0} 
    - \frac{M_zD_z}{\varepsilon_0}
    + \frac{\Omega D_z^2}{2\varepsilon_0}.
\end{equation}
The first two terms form the energy at zero displacement field; $M_z^2/(2\varepsilon_0\Omega)$ is the dipole-correction energy~\citep{yeh_ewald_1999}. The last term is the field energy of the empty cell; this term is left out of the energy labels. It does not depend on the atomic positions and therefore does not contribute to the forces. Leaving it out, as in the finite-field \gls{dft} energy below, makes $-\partial U/\partial E_0$ equal to the dipole $M_z$ of the atoms and electrons, without the empty-cell term $\varepsilon_0\Omega E_0$.

$U_0$ and $M_z$ also depend on the electronic state $\psi$. A finite-field \gls{dft} calculation returns 
\begin{equation}
  U(\mathbf{R},E_0) = \min_\psi\bigl[U_0(\psi;\mathbf{R}) - M_z(\psi;\mathbf{R})\,E_0\bigr],
\end{equation}
with $\psi^*$ the minimizing state. Because $\psi^*$ is stationary, its dependence on $E_0$ does not enter the first derivative of $U$, so that
\begin{equation}
  \frac{\partial U}{\partial E_0}\bigg|_\mathbf{R} = \frac{\partial U}{\partial \psi}\bigg|_{\psi^*} \frac{\partial \psi^*}{\partial E_0} - M_z(\psi^*;\mathbf{R}) = -M_z(\psi^*;\mathbf{R})
\end{equation}
(Hellmann--Feynman theorem). The \gls{dft} labels therefore satisfy
\begin{equation}
  U(\mathbf{R},E_0) = U_0(\mathbf{R},E_0) - M_z(\mathbf{R},E_0)\,E_0,
  \qquad
  M_z = -\frac{\partial U}{\partial E_0}\bigg|_\mathbf{R},
  \label{eq:sm-applied-field}
\end{equation}
where both $U_0$ and $M_z$ are evaluated at $\psi^*(E_0)$ and thus depend on the field through the electronic polarization.

% \subsection{When the field direction is required} \label{sm:symmetry}
% \begin{plan}
%   \item One water between Ne layers (\texttt{water\_Ne\_single}): scalar-conditioned arms give the $\cos\theta$ coefficient of $M_z$ exactly zero; vector-coupled arms reproduce DFT (0.3675\,e\AA). Table.
%   \item Field-parity test on a symmetric water slab (\texttt{water\_slab\_field\_parity}): an invariant model passes the standard test set and fails on rotated copies whose labels follow from symmetry.
%   \item Open: the lateral-image $\cos2\theta$ term.
% \end{plan}

\subsection{Field-conditioned LOREM models} \label{sm:field}
We implemented field conditioning in LOREM~\citep{rumiantsev_learning_2026} following MACE-Field~\cite{martin_general_2026}. 
We contract the tensorial features $\mathbf{h}_i^{(l)}$ with the applied field in a weighted tensor product,
\begin{equation}
    {\mathbf{h}}^{(l)}_i \leftarrow  \mathbf{h}_i^{(l)} - \mathbf{h}^{(l)}_i \otimes_\mathbf{w} \mathbf{E}_0^{(1)},
\end{equation}
where $\mathbf{E}^{(1)}_0$ is the $l=1$ spherical harmonic representation of the applied field $E_0\hat{\mathbf{z}}$. The energy is constructed from the last-layer scalar features as $U=\sum_iU_i({\mathbf{h}}^{(0)}_i)$. Because the cell dipole $M_z=-\partial U/\partial E_0$ is obtained from automatic differentiation, the model effectively represents it as a sum over atomic dipole contributions, $M_z=\sum_i \mu_{i,z}$.

For a consistent comparison, the field-conditioned long-range and short-range models are the same LOREM model and differ only in the long-range term.
Both use 64 scalar and 8 spherical features up to $l=2$, 16 radial basis functions, a \qty{5}{\angstrom} cutoff and one message-passing step.
The long-range model uses an Ewald sum over the latent charges ($l=0$, scalar only) with a slab correction along $z$, which is the boundary condition of the reference calculations.
The short-range model keeps only its \qty{10}{\angstrom} effective cutoff.
We train on energy, forces and $M_z$ with loss weights $1000:1:0.5$, using the Muon optimizer~\citep{liu_muon_2025} with a learning rate of $2\times10^{-4}$, 20 warm-up epochs, cosine decay, gradient clipping at 1.0, batches of 7 structures and 400 epochs.
We keep the model from the training step with the highest summed $R^2$ of energy, forces and dipole on the validation set (early stopping).
We do not train on the Born effective charge here.

\subsection{Training and simulation} \label{sm:ne-data}
Structures were obtained from an \gls{aimd} trajectory from Ref.~\citenum{yang_macro-dipole-constrained_2026}. 
The cell measures $8.464\times9.773\times\qty{44.06}{\angstrom^3}$ and holds 64 H$_2$O between two Ne layers, 216 atoms in total, with about \qty{22}{\angstrom} of water.
It is periodic in $x$ and $y$ and non-periodic along $z$.
To label the data we run \gls{dft} calculations with VASP~6.4.3~\citep{kresse1993ab, kresse1994ab, kresse1996efficiency, kresse1996efficient, kresse1999ultrasoft} in a finite field $E_0\hat{\mathbf{z}}$ with dipole correction. Other than that, the settings are the same as the reference \gls{aimd}~\citep{yang_macro-dipole-constrained_2026}: RPBE \citep{hammer1999improved} with D3 dispersion \citep{grimme2010consistent} (zero damping), a \qty{500}{\electronvolt} plane-wave cutoff, a $2\times2\times1$ Monkhorst--Pack~\citep{monkhorst1976special} grid, \qty{0.1}{\electronvolt} Fermi smearing and an electronic convergence of \qty{e-6}{\electronvolt}. To avoid dielectric breakdown we discard labels with a Kohn--Sham gap below \qty{1.5}{\electronvolt} or a smearing entropy above \qty{1}{\milli\electronvolt}.

To generate the dataset, we first sample 1146 structures from the zero-field trajectory of Ref.~\citenum{yang_macro-dipole-constrained_2026}. Of these, 712 enter the training set with their zero-field labels only. The remaining 434 are labeled under an applied field: 302 for training, 44 for validation and 88 for testing. Training and validation frames are labeled with a pair of positive and negative fields, $\pm E_0$ with $|E_0|$ drawn uniformly from $[0.05,0.2]$~\unit{\volt/\angstrom}. Test frames get $\pm0.1$ and $\pm 0.2$ \unit{\volt/\angstrom}. This gives 712 plus $3\times302$ (two fields and the original zero-field label) training structures. Five test structures at $\pm0.2$\unit{\volt/\angstrom} did not pass the breakdown test and were left out. This constitutes round 0 (see Table~\ref{tab:sm-ne-data}).

We then grow the dataset in two active-learning rounds. We run molecular dynamics (settings are given below) at $E_0=0,\pm0.1,\pm0.2$ \unit{\volt/\angstrom} for \qty{110}{ps} (\qty{10}{ps} equilibration, \qty{100}{ps} production), and sample 50 frames from each run (one every 2 ps) for both models, resulting in 500 additional geometries per round. These 500 frames were split evenly into 400 training, 50 validation and 50 test frames. Training and validation frames were labeled once at $E_0+\delta$, with $\delta$ normally distributed with width \qty{0.05}{\volt/\angstrom} truncated at $|\delta|\leq\qty{0.08}{\volt/\angstrom}$; test frames were labeled at the pair $E_0\pm\qty{0.05}{\volt/\angstrom}$. In round 1, 104 of 110 labels from the short-range model trajectories at $E_0\pm\qty{0.2}{\volt/\angstrom}$ had to be discarded because the overpolarization led to dielectric breakdown.

\begin{table}[h]
\caption{The Ne--water--Ne dataset per round. Training and validation sets are cumulative; each round has its own test set. Round 0's training set includes the 712 zero-field-only geometries.}
\label{tab:sm-ne-data}
\begin{ruledtabular}
\begin{tabular}{llrrrrr}
Round & Source & Geometries & Kept labels & Training & Validation & Test \\
\hline
0 & AIMD, every 45th step & 434 & 1473 & $906+712=1618$ & 132 & 435 \\
1 & round-0 MD, both models & 500 & 398 & 1905 & 168 & 75 \\
2 & round-1 MD, both models & 500 & 529 & 2291 & 215 & 96 \\
\end{tabular}
\end{ruledtabular}
\end{table}

The test set root-mean-squared errors (RMSEs) are shown in Table~\ref{tab:sm-ne-errors}. In round 0, both models achieve similar errors, with the long-range model performing slightly better. In rounds 1 and 2, the short-range errors are about twice as large in energy and 1.5 times as large in forces, while its dipole errors stay close to those of the long-range model.
The difference in the sampled dipoles (Fig.~\ref{fig:dipole}) is therefore much larger than any difference in these test errors.
\begin{table}[h]
\caption{Test RMSEs of the round-2 models on the test set of each round. Energies are per structure (216 atoms).}
\label{tab:sm-ne-errors}
\begin{ruledtabular}
\begin{tabular}{llrrr}
Model & Test set & $E$ (meV) & $F$ (meV/\AA) & $M_z$ ($m e$\AA) \\
\hline
long-range & round 0 & 34.8 & 17.6 & 24.7 \\
          & round 1   & 49.2 & 19.9 & 26.9 \\
          & round 2   & 42.9 & 19.3 & 25.9 \\
short-range & round 0 & 41.4 & 21.1 & 30.5 \\
          & round 1   & 86.8 & 30.5 & 33.4 \\
          & round 2   & 108.4 & 29.6 & 35.4 \\
\end{tabular}
\end{ruledtabular}
\end{table}

We run molecular dynamics using the Atomic Simulation Environment~\citep{hjorth2017atomic} at constant $E_0$ with a CSVR thermostat \citep{bussi2007canonical} (\qty{100}{fs} relaxation time) at \qty{300}{\kelvin} and a \qty{0.5}{\femto\second} time step, keeping the Ne atoms fixed.
In the \qty{22}{\angstrom} cell, we run the short-range and long-range models of every round at $E_0=0$, $\pm0.1$ and $\pm\qty{0.2}{\volt\per\angstrom}$ for \qty{10}{\pico\second} of equilibration and \qty{100}{\pico\second} of production.
The \qty{31}{\angstrom} cell holds 83 H$_2$O in a \qty{34}{\angstrom} gap between the Ne layers ($L_\mathrm{w}=\qty{31.34}{\angstrom}$).
We run the round-2 models in this cell at $E_0=0$ and $\pm\qty{0.1}{\volt\per\angstrom}$ for \qty{50}{\pico\second} of equilibration and \qty{150}{\pico\second} of production.
No models are trained on this cell.
We compute $\langle M_z\rangle$ and $\langle\cos\theta\rangle(z)$ from the production runs, with $\langle\cos\theta\rangle(z)$ on \qty{1}{\angstrom} bins and leaving out bins that hold less than half a molecule per frame on average.
Error bars are block estimates over ten blocks.

\subsection{Dipole response from fluctuations} \label{sm:fluctuations}
In terms of the energy in Eq.~\eqref{eq:sm-applied-field}, the mean cell dipole is
\begin{equation}
  \langle M_z\rangle = \frac{1}{Z}\int M_z(\mathbf{R},E_0)\,e^{-\beta U(\mathbf{R},E_0)}\,\mathrm{d}\mathbf{R},
  \qquad
  Z = \int e^{-\beta U(\mathbf{R},E_0)}\,\mathrm{d}\mathbf{R},
\end{equation}
with $\beta=1/k_\mathrm{B}T$.
Differentiating with respect to $E_0$ and using Eq.~\eqref{eq:sm-applied-field}, we obtain 
\begin{align}
  \frac{\partial\langle M_z\rangle}{\partial E_0}
  &= \biggl\langle\frac{\partial M_z}{\partial E_0}\bigg|_\mathbf{R}\biggr\rangle
   - \beta\biggl\langle M_z\frac{\partial U}{\partial E_0}\biggr\rangle
   + \beta\langle M_z\rangle\biggl\langle\frac{\partial U}{\partial E_0}\biggr\rangle \nonumber\\
  &= \langle\alpha_{zz}\rangle + \beta\bigl(\langle M_z^2\rangle - \langle M_z\rangle^2\bigr).
  \label{eq:sm-fluctuation-general}
\end{align}
With $\alpha_{zz}=\partial M_z/\partial E_0|_\mathbf{R}$, this is Eq.~\eqref{eq:fluctuation} of the main text. The first term is the electronic response at frozen nuclei and the second (the variance) the orientational response of the nuclei, which follows from the equilibrium fluctuations of the dipole.
Macroscopic electrostatics gives a closed-form estimate for the bias response. At fixed $\mathbf{D}=\varepsilon_0\mathbf{E} + \mathbf{P}$, linear response gives $\mathbf{D}=\varepsilon\mathbf{E}$, so that $\mathbf{P}=(1-1/\varepsilon)\mathbf{D}$. The dipole moment is $\mathbf{M}=AL_\mathrm{w}\mathbf{P}$, with $A$ the area and $L_\mathrm{w}$ the thickness of the water slab. With $\mathbf{D}=\varepsilon_0E_0\hat{\mathbf{z}}$, 
\begin{equation}
  \frac{\partial\langle M_z\rangle}{\partial E_0} = \varepsilon_0 A L_\mathrm{w}\left(1-\frac{1}{\varepsilon}\right).
  \label{eq:sm-continuum}
\end{equation}
Taking the derivative at constant $\mathbf{R}$ (frozen nuclei) leaves only the electronic response, so that $\varepsilon$ is replaced by the high-frequency permittivity $\varepsilon_\infty$. 
Inserting these estimates in Eq.~\eqref{eq:fluctuation} gives an estimate for the variance,
\begin{equation}
  \text{Var} ( M_z) = \varepsilon_0 A L_\mathrm{w}\, k_\mathrm{B}T\left(\frac{1}{\varepsilon_\infty}-\frac{1}{\varepsilon}\right),
  \label{eq:sm-continuum-variance}
\end{equation}
which is used to plot the dashed Gaussian curve in Fig.~\ref{fig:dipole}b,c. Because $1/\varepsilon$ is comparatively small for water, the choice of water permittivity has little effect; we use $\varepsilon=78.5$. With $\varepsilon_\infty=\num{1.89}$, $L_\mathrm{w}=\qty{22.0}{\angstrom}$ and the \gls{aimd} temperature of \qty{298}{\kelvin}, this gives a standard deviation of \qty{0.365}{}~$e\,\unit{\angstrom}$, close to the \qty{0.373}{}~$e\,\unit{\angstrom}$ of the \gls{aimd} trajectory itself. 

For a more accurate estimate of the bias response, the electronic response at frozen nuclei is evaluated from the test set by central finite differences,
\begin{equation}
  \alpha_{zz}\approx\frac{M_z(+\Delta E)-M_z(-\Delta E)}{2\Delta E}
\end{equation}
with $\Delta E=\qty{0.1}{\volt\per\angstrom}$ (two single-point calculations per snapshot), averaged over snapshots of the zero-field \gls{aimd}. The variance is also obtained from the \gls{aimd} trajectory. For the \qty{22}{\angstrom} film, we obtain \num{5.26} for the frozen-nuclei term and \num{5.42}~$e\,\unit{\angstrom}$ per \unit{\volt\per\angstrom} for the variance term $\beta\,\mathrm{Var}(M_z)$, so $\partial \langle M_z \rangle/\partial E_0|_{E_0=0}$ is \num{10.7} $e\,\unit{\angstrom}$ per \unit{\volt\per\angstrom}, the slope of the dashed line in Fig.~\ref{fig:dipole}d,e.

\subsection{Reference for the thick film} \label{sm:continuum}
Eq.~\eqref{eq:sm-continuum} also gives an estimate for the response of the thick film. With $L_\mathrm{w}=\qty{31.34}{\angstrom}$ we obtain an estimate of \num{14.15}~$e$\unit{\angstrom} per \unit{\volt\per\angstrom}.
However, with $L_\mathrm{w}=\qty{22.0}{\angstrom}$, Eq.~\eqref{eq:sm-continuum} gives \num{9.93}~$e$\unit{\angstrom} per \unit{\volt\per\angstrom} for the \qty{22}{\angstrom} cell, \qty{7}{\percent} below the value \num{10.7} obtained from DFT and \gls{aimd} through Eq.~\eqref{eq:fluctuation}.
The response of interfacial water likely differs from that of a macroscopic continuum.
Instead, we therefore take the response from the \qty{22}{\angstrom} cell and add the response of the additional (bulk-like) water:
\begin{equation}
  \frac{\partial\langle M_z\rangle}{\partial E_0}\bigg|_{\qty{31}{\angstrom}}
  = \frac{\partial\langle M_z\rangle}{\partial E_0}\bigg|_{\qty{22}{\angstrom}} + \varepsilon_0 A\,\Delta L_\mathrm{w}\left(1-\frac{1}{\varepsilon}\right),
\end{equation}
with $\Delta L_\mathrm{w}=\qty{9.34}{\angstrom}$ the difference between the water thicknesses of the two cells, which gives \num{14.90}~$e$\unit{\angstrom} per \unit{\volt\per\angstrom}, the slope of the dashed line in Fig.~\ref{fig:dipole}g. This value lies \qty{5}{\percent} above the continuum estimate.
The round-2 long-range model gives \num{14.29}~$e$\unit{\angstrom} per \unit{\volt\per\angstrom}, between the two estimates, whereas the short-range model gives \num{25.74}~$e\,\unit{\angstrom}$ per \unit{\volt\per\angstrom}.

\subsection{Work function of a metal--water--countercharge cell} \label{sm:workfunction}

\begin{figure}[b]
    \centering
    \includegraphics{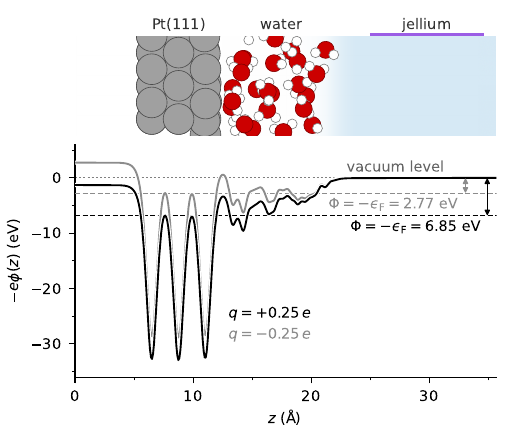}
    \caption{Planar-averaged potential energy of an electron, $-e\phi(z)$, in a Pt(111)/water cell from Ref.~\citenum{bergmann_erasing_2026} with a solvated jellium counter charge (GPAW-SJM), at $q=-0.25$ (gray) and $+0.25e$ (black), with the vacuum level above the jellium ($\psi$) taken as zero. Part of the structure is shown on top and its scale matches the $z$-axis. Dashed lines mark the Fermi level $\epsilon_\mathrm{F}=\mu^\mathrm{m}-e\phi^\mathrm{m}$ for the two different charges. The implicit solvent is indicated by the blue shaded region; the position of the jellium is marked by the purple bar. The step at the cell boundary is the dipole correction.}
    \label{fig:sm-potential}
\end{figure}

In this section we keep the elementary charge $e$ explicit, so that the work function $\Phi$ (or absolute electrode potential) is an energy; the main text uses units with $e=1$.
We consider a metal slab covered by a water film, with an excess charge $q$ on the electrode and a counter charge $-q$ above the water (Fig.~\ref{fig:sm-potential}); $z$ is the surface normal and $A$ the lateral area.
The Fermi level is the electrochemical potential of the electrons, $\epsilon_\mathrm{F}=\mu^\mathrm{m}-e\phi^\mathrm{m}$, where $\mu^\mathrm{m}$ is their chemical potential in the metal and $\phi^\mathrm{m}$ the electrostatic potential inside it.
The work function is the energy to move an electron from the Fermi level to the vacuum just outside the counter charge, where the electrostatic potential is the outer potential $\psi$:
\begin{equation}
  \Phi = -e\psi - \epsilon_\mathrm{F} = -\mu^\mathrm{m} + e\chi, \qquad \chi = \phi^\mathrm{m} - \psi,
  \label{eq:sm-wf}
\end{equation}
with $\chi$ the potential drop from the metal to the vacuum.
Because $q=-e\Delta N_e$ and $\epsilon_\mathrm{F}=\partial U/\partial N_e$, taking $\psi=0$ gives $\partial U/\partial q = -\epsilon_\mathrm{F}/e = \Phi/e$, which is $\Phi=\partial U/\partial q$ of the main text with $e=1$.

To obtain $\chi$, let $\rho(z)=A^{-1}\int\rho(\mathbf{r})\,\mathrm{d}x\,\mathrm{d}y$ be the charge density averaged over the lateral cell, including the electrode charge and the counter charge.
We take $z_\mathrm{m}$ inside the metal and $z_\mathrm{vac}$ in the vacuum above the counter charge. The field vanishes at both points, so by Gauss's law the enclosed charge is zero: $\int_{z_\mathrm{m}}^{z_\mathrm{vac}}\rho(z')dz'=0$. To obtain $\chi=\phi^\mathrm{m}-\psi$, we integrate Poisson's equation twice between $z_\mathrm{m}$ and $z_\mathrm{vac}$
\begin{equation}
    -\chi =\psi - \phi^\mathrm{m} = -\frac{1}{\varepsilon_0} \int_{z_\mathrm{m}}^{z_\mathrm{vac}} dz \int_{z_\mathrm{m}}^z \rho(z')dz' = \frac{1}{\varepsilon_0} \int_{z_\mathrm{m}}^{z_\mathrm{vac}} z \rho(z)dz,
\end{equation}
where we switched the order of integration over the triangular integration domain. The integral is the $z$-component of the dipole moment of the cell per unit area: $\chi = -{M_z}/(\varepsilon_0 A)$.

We compare with a reference system, such as the bare slab at $q=0$, with work function $\Phi_\mathrm{ref}$.
For a slab thick enough that its interior is unperturbed, $\mu^\mathrm{m}$ is the same in both, so $\Phi-\Phi_\mathrm{ref}=e\Delta \chi$, which depends on the charge density difference $\Delta\rho$ with respect to the reference system.
Since $\Delta\rho$ vanishes below $z_\mathrm{m}$, the integral can be extended over the whole cell:
\begin{equation}
  \Phi = \Phi_\mathrm{ref} - \frac{e}{\varepsilon_0 A}\int (z-z_\mathrm{m})\Delta\rho(\mathbf{r})d^3\mathbf{r}.
  \label{eq:sm-wf-drho}
\end{equation}

We partition $\Delta\rho$ into contributions $\Delta\rho_i$ of the atoms and the counter charge $\rho_\mathrm{cc}$, $\Delta\rho=\sum_i\Delta\rho_i+\rho_\mathrm{cc}$.
For atom $i$ at height $z_i$, we define its charge and dipole,
\begin{equation}
  \Delta q_i = \int\Delta\rho_i d^3\mathbf{r}, \qquad
  \Delta\mu_{i,z} = \int (z-z_i)\Delta\rho_i d^3\mathbf{r},
\end{equation}
with $\sum_i\Delta q_i = q = -\int\rho_\mathrm{cc} d^3\mathbf{r}$.
Splitting $z-z_\mathrm{m}=(z-z_i)+(z_i-z_\mathrm{m})$ for each atom, and introducing the centroids of the interfacial charge and of the counter charge,
\begin{equation}
  \bar z_q = \frac{1}{q}\sum_i z_i\,\Delta q_i, \qquad
  \bar z_\mathrm{cc} = -\frac{1}{q}\int z \rho_\mathrm{cc} d^3\mathbf{r},
\end{equation}
the integral in Eq.~\eqref{eq:sm-wf-drho} becomes
\begin{equation}
  \int (z-z_\mathrm{m})\Delta\rho d^3\mathbf{r}
  = \sum_i\bigl[\Delta\mu_{i,z} + (z_i-z_\mathrm{m})\Delta q_i\bigr] - q(\bar z_\mathrm{cc}-z_\mathrm{m})
  = \Delta M_z - q d,
\end{equation}
with $\Delta M_z=\sum_i\Delta\mu_{i,z}$ and $d=\bar z_\mathrm{cc}-\bar z_q$; the terms in $z_\mathrm{m}$ cancel because the electrode charge and the counter charge are equal and opposite.
Hence
\begin{equation}
  \Phi = \Phi_\mathrm{ref} - \frac{e\Delta M_z}{\varepsilon_0 A} + \frac{eq d}{\varepsilon_0 A},
\end{equation}
which is Eq.~\eqref{eq:work-function-decomposition} of the main text with $e=1$.
The atomic dipoles $\Delta\mu_{i,z}$ contain the electronic rearrangement of both the metal surface and the water. 
The partition into atomic contributions is not unique; an MLIP that predicts $\Phi$ from a sum over atoms learns one such partition.

\section{Metal--water interfaces} \label{sm:metal}

\subsection{Charge conditioning} \label{sm:charge}
We implement charge conditioning in LOREM and GRACE using \gls{film}~\citep{perez_film_2018}, which transforms the descriptor features using two small neural networks $\boldsymbol{\gamma}$ and $\boldsymbol{\beta}$ that take the conditioning parameter $q$ or $\sigma=q/A$ as input,
\begin{equation}
    {\mathbf{h}}_i \leftarrow (\mathbf{1}+\boldsymbol{\gamma}(q)) \odot \mathbf{h}_i + \boldsymbol{\beta}(q),
\end{equation}
where $\odot$ indicates element-wise multiplication. The work function $\Phi=\partial U/\partial q$ is obtained by automatic differentiation in the same backward pass as the forces. The $z$ column of the Born effective charge tensors, $\mathbf{Z}^*_{i,z} = \varepsilon_0 A\, \partial \mathbf{F}_i/\partial q$, requires an additional backward pass.
For the \qty{108}{atom} Pt(111)/water cell and a \qty{0.5}{\femto\second} time step, with the work function written every 100 steps, our \loremq{} production runs with ASE reach \qty{3.0}{\nano\second/day} (\num{70} steps per second) on one NVIDIA A100.

\subsection{\grace-\razor{}} \label{sm:razor}
The \gls{razor} approach~\cite{bergmann_machine_2025, bergmann_erasing_2026} uses two backbones, one for $U_0$ and one for $\Phi$. We implement the same logic in a single \grace{} backbone, which improves computational efficiency and simplifies the integration in the \gls{lammps} code for molecular dynamics. By forming latent atomic dipoles $\boldsymbol{\mu}_i$ from the $l=1$ last-layer features and contracting them with $\mathbf{D}=(q/A)\hat{\mathbf{z}}$, the atomic contribution to the energy becomes
\begin{equation}
    U_i = U_{0,i} + q\left(\frac{\Phi_\mathrm{ref}}{N} - \frac{\mu_{i,z}}{\varepsilon_0 A}\right) + \frac{1}{2}q^2\,\frac{1}{C_0 N},
\end{equation}
where $N$ is the number of atoms in the cell.
Summed over atoms, this recovers Eq.~\eqref{eq:razor-expansion} of the main text. 
Forces, work function and Born effective charges follow from automatic differentiation.
$\Phi_\mathrm{ref}$ is degenerate with a constant in $\sum_i\mu_{i,z}$; we fix it at the clean Pt(111) value of \qty{5.75}{\volt}, and learn only $C_0^{-1}$, initialized at the mean $\partial^2U/\partial q^2$ of the training set.
The area $A$ is a fixed parameter of the model rather than read from the cell, because the \gls{lammps} interface does not pass the cell to the model.

For the \qty{108}{atom} Pt(111)/water cell and a \qty{0.5}{\femto\second} time step, with the work function written every step, the production runs of this work reach \PL{\qty{5.5}{\nano\second/day}} (\num{127} steps per second) on one NVIDIA A100.

\subsection{One water above Pt(111) and Pt(332)} \label{sm:pt}
The structures consist of one water molecule above a four-layer Pt slab, flat Pt(111) ($9.749\times\qty{8.443}{\angstrom^2}$) or stepped Pt(332) ($13.200\times\qty{8.443}{\angstrom^2}$, one step per cell), in a cell of \qty{34.81}{\angstrom} along $z$.
The water geometries are 20 gas-phase molecules of different dipole magnitude.
For training, we create 1200 frames per surface. For each frame, we draw a random water geometry, give it a uniformly random orientation, and place it at a height $h\in[5,10]$~\unit{\angstrom} above the surface, with the lateral position sampled from a uniform distribution.
We keep the 964 frames with $h\le\qty{9}{\angstrom}$: for higher waters, the highest occupied orbital approaches the Fermi level at positive charge, so that electrons start to leave the water, an effect not expected in a typical simulation setup.
We split them at random into 868 training and 96 validation frames per surface.
For the test curves, we choose a water geometry with median dipole; on Pt(111), its dipole is rotated by the polar angle $\theta$ at fixed position; on Pt(332), the water is moved by $x$ across the step at fixed dipole.

We label the dataset with ABACUS 3.10.1~\cite{zhou_abacus_2025} in its LCAO mode, choosing settings that provide internal consistency but not necessarily absolute accuracy. 
All models are trained and tested on the same labels, so the comparison between the architectures does not depend on the accuracy of these settings; only the absolute work functions are not converged.
We use PBE~\citep{perdew1996generalized}, no dispersion, SG15 ONCV~\citep{schlipf2015optimization, hamann2013optimized} pseudopotentials, DZP orbitals (Pt and O \qty{7}{bohr}, H \qty{6}{bohr}), a \qty{100}{Ry} grid, Gaussian smearing of \qty{0.02}{Ry}, and Monkhorst--Pack~\citep{monkhorst1976special} grids of $4\times4\times1$ (Pt(111)) and $3\times4\times1$ (Pt(332)).
Each frame carries a charge $q$ drawn uniformly from $[-0.030, +0.025]$~$e$ on Pt(111). For Pt(332), we use the same surface-charge density range.
The counter charge $-q$ sits on a plate \qty{2.5}{\angstrom} above the highest atom of all structures of a surface, behind which a barrier of \qty{1.5}{Ry} keeps the electrons out of the vacuum; the plate and barrier are fixed in space.
The work function is read from the planar-averaged potential in the vacuum beyond the plate.
On a charge stencil, $\partial U/\partial q$ from the energies matches this work function to \qty{0.5}{\milli\volt}, so the label $\Phi$ is the charge derivative of the energy label.
The small basis keeps each calculation at 2--3 minutes on one node; it shifts the work function by a constant and changes the step effect by about \qty{5}{\percent}, so the labels are consistent rather than converged.

\grace{} models use two layers with a \qty{6}{\angstrom} cutoff; \lorem{} uses a \qty{5}{\angstrom} cutoff with one message-passing step plus the long-range Ewald term.
All are trained on energy, forces and work function.
Table~\ref{tab:sm-pt-errors} lists the validation errors and the errors on the curves of Fig.~\ref{fig:pt}.
\grace-\razor{} and \loremq{} fit the validation energies and forces equally well, but on Pt(332) the work-function error of \grace-\razor{} is four times larger, and only \loremq{} follows the step at both heights.

\begin{table}[h]
\caption{Errors of the models in Fig.~\ref{fig:pt}: RMSE on the 96 validation frames per surface, and RMSE of $\Phi$ on the curves drawn in Fig.~\ref{fig:pt}, at $h=5$ and \qty{7}{\angstrom} (inside and beyond the \grace{} cutoff).}
\label{tab:sm-pt-errors}
\begin{ruledtabular}
\begin{tabular}{llrrrrr}
 & & \multicolumn{3}{c}{validation} & \multicolumn{2}{c}{curves, $\Phi$ (mV)} \\
surface & model & $E$ (meV/atom) & $F$ (meV/\AA) & $\Phi$ (mV) & $h=5$ & $h=7$ \\
\hline
Pt(111) & \loremq & 0.17 & 3.5 & 1.9 & 3.7 & 1.2 \\
        & \grace-\razor & 0.14 & 3.1 & 2.9 & 2.5 & 3.2 \\
        & \graceq & 0.52 & 46.6 & 470 & 76 & 666 \\
Pt(332) & \loremq & 0.10 & 2.4 & 1.8 & 3.1 & 0.9 \\
        & \grace-\razor & 0.10 & 2.5 & 7.7 & 16.4 & 9.9 \\
        & \graceq & 0.40 & 40.3 & 386 & 432 & 667 \\
\end{tabular}
\end{ruledtabular}
\end{table}

\section{Pt(111)--water} \label{sm:md}

% \begin{plan}
%   \item Data (\razor\ publication split; labels as $\partial U/\partial Q$), test errors against \razor-qQ.
%   \item MD protocol (\qty{350}{\kelvin}, CSVR, four replicas of \qty{250}{\pico\second}), AIMD reference (GPAW-SJM, PBE-D3, two chains per charge), sampling and error bars.
%   \item \textbf{Work-function label (must be discussed):} all Pt(111) models are trained on the SJM work function $W$, which lies $\approx\qty{0.3}{\volt}$ above $\partial U/\partial q$ of the \gls{dft} energy labels. The labels are therefore not exactly consistent; the models learn $\Phi=\partial U/\partial q$ and match $W$. Show both scales and the effect on $\langle\Phi\rangle(q)$ (\texttt{lorem\_dEdq}, \texttt{grace\_dEdq} arms).
%   \item The potential of the \grace{} single-$q$ series is scored with \loremq{}.
%   \item $\langle\Phi\rangle(Q)$, capacitance and Eq.~\eqref{eq:capacitance}; significance of the AIMD peak.
%   \item Chemisorption at negative charge; equilibration of the published AIMD.
%   \item Active-learning procedure for the \grace{} single-$q$ reference models (rounds, selection criterion, number of added structures).
%   \item Exclusion of $q=+1\,e$ from training: evidence for dielectric breakdown in these data~\cite{bergmann_erasing_2026}.
% \end{plan}

\subsection{Training data}
The Pt(111)--water models are trained on the published data of Ref.~\citenum{bergmann_erasing_2026}: a three-layer $3\times4$ Pt(111) slab (36 Pt) with 24 water molecules, lateral area \qty{82.31}{\angstrom^2}, at bias charges $|q|\le1\,e$, computed with GPAW and the solvated jellium method (SJM)~\cite{mortensen2024gpaw, kastlunger_controlled-potential_2018} using PBE-D3(BJ)~\citep{perdew1996generalized, grimme2010consistent, grimme_effect_2011}, $\Gamma$-point only. 

For this system, the charge response is only linear within a window of about $0.5e$ around the charge it was equilibrated at (Fig.~\ref{fig:sm-labels}a). 
Outside that window, electrons are deposited at or removed from states of the water (dielectric breakdown) \citep{bergmann_erasing_2026}. 
For structures equilibrated at $q_\mathrm{MD}=+1e$, the window is centered near $+0.75e$, and at $+1e$ they already sit on the upper end of the linear region.
For the models and results described in the main text, we therefore train and run MD only up to $q\le+0.75e$.

\subsection{Work-function labels} \label{sm:labels}
The data contain two work function labels for each structure.
$\Phi_\mathrm{SJM}$ is the work function returned by GPAW-SJM.
$\Phi_\mathrm{fd}$ is $\partial U/\partial q$ obtained by finite differences of the energy on a charge stencil of $\pm0.25e$, which Ref.~\citenum{bergmann_erasing_2026} uses for its labels.
For an energy surface $U(\mathbf{R},q)$, both should be equal, but they differ by $\delta=\Phi_\mathrm{SJM}-\Phi_\mathrm{fd}\approx\qty{0.3}{\volt}$. A model with $\Phi=\partial U/\partial q$ trained on $\Phi_\mathrm{SJM}$ and on the SJM energies therefore receives labels that are not exactly consistent.

The difference is not an artifact of the finite-difference derivative.
On 28 AIMD-sampled structures at $q=-1$ to $+1\,e$, we recomputed $\partial U/\partial q$ with a stencil of $\pm0.05\,e$ and obtained $\delta=0.28\pm0.09\unit{\volt}$ (standard deviation over structures), against $0.27\pm0.08\unit{\volt}$ from the $\pm0.25e$ labels of the same structures. We have not identified the origin of $\delta$; it implies that a term in the GPAW-SJM energy is absent from the Kohn--Sham potential that determines the Fermi level. The per-charge means of $\delta$ change by only $0.03\pm0.02\unit{\volt}$ per $e$.

\begin{figure}
    \centering
    \includegraphics{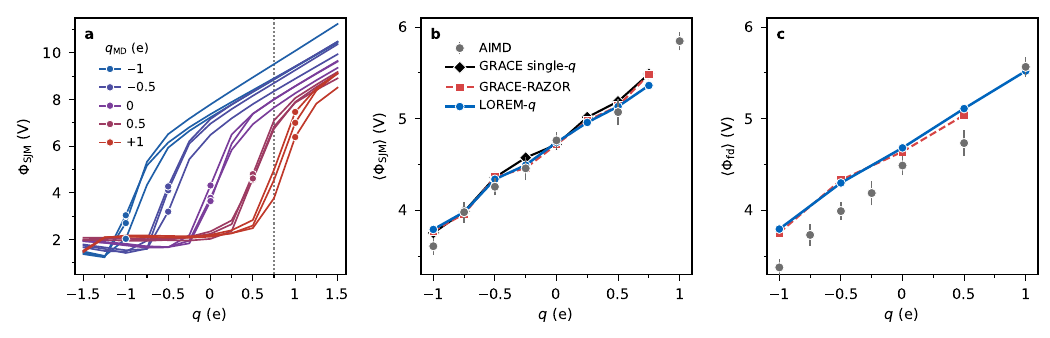}
    \caption{Work function of the Pt(111)--water models.
    (a) $\Phi_\mathrm{SJM}$ of single structures from Ref.~\citenum{bergmann_erasing_2026}, each equilibrated by \gls{md} at a charge $q_\mathrm{MD}$ (dots) and recomputed with GPAW-SJM at charges from $-1.5$ to $+1.5e$. The dotted line marks the highest training charge, $+0.75\,e$. 
    (b) $\langle\Phi_\mathrm{SJM}\rangle(q)$ from \gls{md} with the models trained on $\Phi_\mathrm{SJM}$ and from AIMD---the same as Fig.~\ref{fig:capacitance}h. The work function of the \grace{} single-$q$ series is computed with \loremq{}.
    (c) $\langle\Phi_\mathrm{fd}\rangle(q)$ from \gls{md} (here, two replicas only) with the same architectures trained on $\Phi_\mathrm{fd}$ (same symbols as in b), with the AIMD converted to $\Phi_\mathrm{fd}$ by subtracting the average offset $\delta(q)=\Phi_\mathrm{SJM}-\Phi_\mathrm{fd}$ (Sec.~\ref{sm:labels}).
    At $+1\,e$, both \grace-\razor{} replicas and one \loremq{} replica dissociate the water and are excluded.
    Error bars: standard errors over replicas; for the AIMD, as described in Sec.~\ref{sm:mdruns}.}
    \label{fig:sm-labels}
\end{figure}

We train the models of the main text on $\Phi_\mathrm{SJM}$, the quantity reported by the AIMD reference.
Figure~\ref{fig:sm-labels}b,c compares them with the same architectures trained on $\Phi_\mathrm{fd}$.
For the comparison in panel c, the AIMD is converted to $\Phi_\mathrm{fd}$ by subtracting the measured $\delta(q)$.
Trained on $\Phi_\mathrm{SJM}$, both models follow the AIMD within \qty{0.1}{\volt}, as shown in the main text (Fig.~\ref{fig:capacitance}).
Trained on $\Phi_\mathrm{fd}$, both lie 0.15--\qty{0.3}{\volt} above the converted AIMD; since the two models agree closely with each other, this reflects an inconsistency in the data labels rather than in the architecture.
% Panel c further shows that RAZOR-qQ also lies above the AIMD reference, suggesting that it may suffer from the same issue.

\subsection{Test errors}
Table~\ref{tab:sm-md-errors} compares the models on the 395 test structures with $q\le+0.75e$ of Ref.~\citenum{bergmann_erasing_2026}.
% The \razor-qQ model was re-evaluated to compare total energies and forces, not energies and forces at zero charge, which one of the \razor{} backbones predicts.
% Both \grace-\razor{} and \loremq{} are more accurate than \razor-qQ in energy and forces. 
The \loremq{} model most accurately predicts the work function and Born effective charges.

\begin{table}[h]
\caption{RMSE on the 395 structures with $q\le+0.75\,e$ of the test set of Ref.~\citenum{bergmann_erasing_2026}. $\Phi$ errors shown against both labels, with the label each model was trained on in bold. $Z^*$: the Born effective charges along the surface normal, $Z^*_{i\alpha}=A\varepsilon_0\,\partial F_{i\alpha}/\partial q$, per component over all atoms.}
\label{tab:sm-md-errors}
\begin{ruledtabular}
\begin{tabular}{lrrrrr}
model, $\Phi$ label & $E$ (meV/atom) & $F$ (meV/\AA) & $\Phi_\mathrm{fd}$ (V) & $\Phi_\mathrm{SJM}$ (V) & $Z^*$ ($e$) \\
\hline
\loremq, $\Phi_\mathrm{SJM}$       & 0.55 & 25.5 & 0.315 & \textbf{0.071} & 0.014 \\
\grace-\razor, $\Phi_\mathrm{SJM}$ & 0.47 & 19.2 & 0.321 & \textbf{0.086} & 0.019 \\
\loremq, $\Phi_\mathrm{fd}$        & 0.54 & 25.5 & \textbf{0.086} & 0.317 & 0.015 \\
\grace-\razor, $\Phi_\mathrm{fd}$  & 0.44 & 18.9 & \textbf{0.095} & 0.313 & 0.019 \\
% \razor-qQ~\cite{bergmann_erasing_2026}, $\Phi_\mathrm{fd}$ & 1.16 & 39.2 & \textbf{0.139} & 0.325 & 0.017 \\
\end{tabular}
\end{ruledtabular}
\end{table}

\subsection{Molecular dynamics} \label{sm:mdruns}
The \gls{md} runs use the cell of the training data, with the two bottom Pt layers fixed, periodic boundary conditions in $x$ and $y$, and the harmonic wall on O of Ref.~\citenum{bergmann_erasing_2026} acting from \qty{21}{\angstrom}.
They run at \qty{350}{\kelvin} with a CSVR thermostat \citep{bussi2007canonical} (\qty{100}{\femto\second} relaxation time) and a \qty{0.5}{\femto\second} time step: four replicas of \qty{250}{\pico\second} per charge, each started from a frame of the published \razor-qQ trajectory at that charge.
\grace{} runs in LAMMPS and \loremq{} in ASE.
We discard the first \qty{30}{\pico\second} of each run and keep a frame every \qty{100}{\femto\second}.

The AIMD reference uses GPAW-SJM with the settings of the training data and D3(BJ) forces, a time step of \qty{1}{\femto\second} and the same thermostat settings as described above.
At each charge, we ran two \qty{7}{\pico\second} trajectories, started from frames of two independent MLIP trajectories, and discard the first \qty{2}{\pico\second} of each.

Error bars of the MLIP series are standard errors over the replicas; for the AIMD, with two chains per charge, we use a block standard error scaled by the factor that makes block errors match the scatter between independent MLIP runs (1.71), or half the drift between the two halves of a chain, whichever is larger.

\subsection{\grace{} single-$q$ reference} \label{sm:singleq}
The single-$q$ reference consists of nine standard two-layer \grace{} models (cutoff \qty{6}{\angstrom}, effective cutoff at most \qty{12}{\angstrom}), one per charge from $-1$ to $+1\,e$ in steps of $0.25\,e$, trained on energy and forces only.
Each is trained on \numrange{1000}{1375} structures at its own charge, about half of them from Ref.~\citenum{bergmann_erasing_2026} and the rest new GPAW-SJM labels with the same settings as the original dadta.

The new structures come from our \gls{md} at \qty{400}{K} with previous charge-conditioned models at that charge, sampled every \qtyrange{0.5}{1}{\pico\second} to create a pool of candidates.
From these we select frames with the maximum set coverage (MSC) algorithm~\citep{yu2026maximizing} of the QUESTS package~\citep{schwalbekoda2025information}, applied to SOAP descriptors~\citep{bartok2013representing} of the interfacial O atoms computed with Featomic~\citep{bigi_metatensor_2026}, and starting from the existing labels at that charge, to choose O environments least represented in the dataset.
The descriptor vectors are normalized to have norm 1, and the kernel bandwidth parameter in QUESTS is set to \num{0.085} so that the overlap between physisorbed and chemisorbed O's is about \num{0.1}.
Frames with a work function outside $2.5<\Phi<\qty{7}{\volt}$ are excluded from training, as they likely exhibit dielectric breakdown.
\gls{md} with the first single-$q$ models showed adsorbed OH or O at negative charges, which is unphysical.
We therefore ran one additional round of active learning, sampling the single-$q$ trajectories in the same way.

The production single-$q$ \gls{md} was run with the same settings as the charge-conditioned \graceq{} models (Sec.~\ref{sm:mdruns}), with four \qty{250}{ps} replicas per charge. As these models have no work-function output, the potential of the single-$q$ series is computed with \loremq{} on its frames.

\end{document}